\documentclass[floatfix,twocolumn,showpacs,preprintnumbers,amsmath,amssymb,pra,superscriptaddress,longbibliography]{revtex4-1}
\usepackage{color}
\usepackage[usenames,dvipsnames,svgnames,table]{xcolor}
\usepackage[colorlinks=true,linkcolor=blue,urlcolor=blue,citecolor=blue]{hyperref}

\usepackage{mathtools}
\usepackage{graphicx}
\usepackage{dcolumn}
\usepackage{array}
\usepackage{lipsum}
\usepackage{bm}
\usepackage{subfigure}
\usepackage{amssymb}
\usepackage{multirow}
\usepackage{tabularx}
\usepackage{amsmath}
\usepackage{braket}

\usepackage{centernot}

\renewcommand{\vec}[1]{\mathbf{#1}}

\begin{document}

%\title{Direct Perturbation Method for Testing Kinetic Energy Functionals:\\
%Nonlocal and Laplacian level approximations}
%MP
\title{Imposing Correct Jellium Response Is Key to Predict the Density Response by Orbital-Free DFT}
% \title{Performance of nonlocal and Laplacian-level Kinetic Energy Functionals Under Harmonic Perturbations}
% \title{ Direct perturbation method for testing functionals in Orbital--Free Density Functional Theory: \\
% nonlocal and Laplacian level approximations}
% \title{ Direct Perturbation Method for Testing  nonlocal and Laplacian Level approximations in Orbital--Free Density Functional Theory}

\author{Zhandos A. Moldabekov}
\email{z.moldabekov@hzdr.de}
\affiliation{Center for Advanced Systems Understanding (CASUS), D-02826 G\"orlitz, Germany}
\affiliation{Helmholtz-Zentrum Dresden-Rossendorf (HZDR), D-01328 Dresden, Germany}

\author{Xuecheng Shao}
%MP
\email{xuecheng.shao@rutgers.edu}
\affiliation{Department of Chemistry, Rutgers University, 73 Warren St., Newark, NJ 07102, USA}

\author{Michele Pavanello}
\email{m.pavanello@rutgers.edu}
\affiliation{Department of Chemistry, Rutgers University, 73 Warren St., Newark, NJ 07102, USA}
\affiliation{Department of Physics, Rutgers University, 101 Warren St., Newark, NJ 07102, USA}

\author{Jan Vorberger}
\affiliation{Helmholtz-Zentrum Dresden-Rossendorf (HZDR), D-01328 Dresden, Germany}

\author{Frank Graziani}
\affiliation{Lawrence Livermore National Laboratory (LLNL), California 94550 Livermore, USA}

\author{Tobias Dornheim}
\affiliation{Center for Advanced Systems Understanding (CASUS), D-02826 G\"orlitz, Germany}
\affiliation{Helmholtz-Zentrum Dresden-Rossendorf (HZDR), D-01328 Dresden, Germany}

%%%%%%%%%%%%%%%%%%%%%%%%%%%%%%%%%%%%%%%%%%%%%%%%%%%%%%%%%%%%%%%%%%%%%%%%%%%%%%%%
% Abstract 
%%%%%%%%%%%%%%%%%%%%%%%%%%%%%%%%%%%%%%%%%%%%%%%%%%%%%%%%%%%%%%%%%%%%%%%%%%%%%%%%
\begin{abstract}
Orbital-free density functional theory (OF-DFT) constitutes a computationally
highly effective tool for modeling  electronic structures of systems ranging from room-temperature materials to warm dense matter. Its accuracy critically depends on the
employed kinetic energy (KE) density functional, which has to be supplied as an external input.
In this work  we consider several nonlocal and Laplacian-level KE functionals and use an external harmonic perturbation to compute the static density response at $T=0$K in the linear and beyond linear response regimes. We test for the satisfaction of exact conditions in the limit of uniform densities and for how approximate KE functionals reproduce the density response of realistic materials (e.g., Al and Si) against the Kohn-Sham DFT reference which employs the exact KE. 
The results illustrate that several functionals violate exact conditions in the UEG limit. %Additionally, we test KE functionals' accuracy to describe density responses beyond the linear regime by gradually increasing the density perturbation amplitude and comparing against Kohn-Sham DFT results which employ the exact non-interacting KE functional. 
We find a strong correlation between the accuracy of the KE functionals in the UEG limit and in the strongly inhomogeneous case. This empirically demonstrates the importance of imposing the limit of UEG response for uniform densities and validates the use of the Lindhard function in the formulation of kernels for nonlocal functionals. This conclusion is substantiated by additional calculations  for bulk Aluminum (Al) with a face-centered cubic (fcc) lattice and Silicon (Si) with an fcc lattice, body-centered cubic (bcc) lattice, and semiconducting crystal diamond (cd) state.  The analysis of fcc Al,  and fcc as well as bcc Si data follows closely the conclusions drawn for the UEG, allowing us to extend our conclusions to realistic systems that are subject to density inhomogeneities induced by ions.  
% Analyzing other classes of KE functionals (e.g., based on the quadratic UEG density response function and the jellium-with-gap model) and other types of systems, such as semiconductors, is left for future studies.
\end{abstract}
%%%%%%%%%%%%%%%%%%%%%%%%%%%%%%%%%%%%%%%%%%%%%%%%%%%%%%%%%%%%%%%%%%%%%%%%%%%%%%%%
%%%%%%%%%%%%%%%%%%%%%%%%%%%%%%%%%%%%%%%%%%%%%%%%%%%%%%%%%%%%%%%%%%%%%%%%%%%%%%%%

\maketitle

%%%%%%%%%%%%%%%%%%%%%%%%%%%%%%%%%%%%%%%%%%%%%%%%%%%%%%%%%%%%%%%%%%%%%%%%%%%%%%%%
%%%%%%%%%%%%%%%%%%%%%%%%%%%%%%%%%%%%%%%%%%%%%%%%%%%%%%%%%%%%%%%%%%%%%%%%%%%%%%%%
\section{Introduction} 

First-principles methods based on the electronic density are often used for simulations of electronic structures in physics and chemistry.
Usually, there is a punishing correlation between the increase in the accuracy of  a method and its computational cost. 
Indeed, there is a computational bottleneck for methods such as  Kohn-Sham density functional theory (KS-DFT) or, more accurately, generalized KS-DFT with hybrid exchange-correlation (XC) functionals, which strongly hinders the simulation of large systems (e.g., with the number of particles of the order of $N\sim10^4$ and more). 
The simulation of large systems is considered to be important for the adequate description of processes involving large numbers of particles such as phase transitions \cite{Karasiev2021}, nucleation \cite{PhysRevB.101.054301}, freezing or melting \cite{Pedersen2016, PhysRevE.102.033205}, collective ion oscillation (normal) modes \cite{Maximilian_Schoerner, zhandos2, Hanno}, and to calculate transport properties such as diffusion and  viscosity \cite{Bethkenhagen_PRE, Daligault_PRL, Sjostrom_PRE}. These properties are often accessed by combining molecular dynamics (MD) simulations of ions with the electron forces computed using a density functional theory based method.

Orbital-free density functional theory (OF-DFT)~\cite{witt_2018} is one of the DFT approaches being actively developed  for the simulation of large systems because of its generally linearly scaling computational cost with respect to the number of particles. OF-DFT has been employed successfully for the description of materials~\cite{witt_2018}, melted metals~\cite{aguado2013orbital}, and even nanoparticles~\cite{shao2021efficient}. 
Additionally, the computational cost of OF-DFT is not sensitive to the variation in temperature since it does not use orbitals. 
This is an important computational advantage of OF-DFT when it comes to the simulation of phenomena at high temperatures \cite{Sjostrom_PRL}, at which thermal KS-DFT simulations require large number of orbitals to correctly capture thermal excitations. Furthermore, being computationally inexpensive, OF-DFT can be used as an intermediate step to accelerate KS-DFT based MD simulations by optimizing an initial ionic configuration \cite{Fiedler_PRR_2022}. Moreover, the OF-DFT approach is one of the main tools for the simulation of near-mesoscopic-scale dynamics taking into account the quantum nonlocality in plasmonics \cite{Mortensen_qhd, Plasmonics_PhysRevX}, quantum plasmas \cite{Manfredi_2008_prb, zhandos_pop18, Graziani_CPP_2022, Moldabekov_SciPost_2022} as well as plasmonics in nanomaterials ~\cite{della2022orbital,baghramyan2021laplacian,covington2021coupled}.  Recent advances have extended the applicability of OF-DFT to dynamic structure factors~\cite{white2013orbital} and optical properties of clusters and nanostructures~\cite{jiang2021nonlocal,jiang2021time,jiang2022efficient}.

In contrast to KS-DFT, the non-interacting kinetic energy (KE) functional in OF-DFT is not expressed as a pure functional of the KS orbitals but it is instead expressed as a pure functional of the electron density. The functional is known to exist, but its analytical pure dependency on the electron density is not known exactly and has to be approximated. Therefore, the key problem  is to design a KE functional, and the corresponding non-interacting free energy at finite temperatures.
Following early seminal works by, e.g., Thomas and Fermi (TF) \cite{Fermi, thomas_1927},  Kohn and Sham~\cite{KS_1965}, Perrot \cite{Perrot}, and Wang and Teter (WT) \cite{WT_paper}, it became a standard to design KE functionals that reproduce the correct density response function in the limit of the ideal uniform electron gas (UEG)~\cite{POP_review,review,quantum_theory, zmcpp2017,wang1999orbital}.  This connection to the archetypical free-electron system delivers a certain level of universality, which, e.g., can lead to accurate results for bulk properties of metals and semiconductors \cite{ke_test_jctc, MGP, PGSL,shao2021efficient,huang2010nonlocal,LMGP}.

Using the UEG model, a commonly used relation (constraint) for the construction of a KE functional $T_{\rm s}[n]$ is based on the Lindhard function $\chi_{\rm Lin}$: \cite{KS_1965,wang2002orbital,zhandos_pop18}
\begin{equation}
\label{eq:start}
-\mathfrak{F}\left[\left.\frac{\delta^2 T_{\rm s}}{\delta n(\vec{r})\delta n(\vec{r^{\prime}})}\right|_{\rm n=n_0}\right]=
%-\left[
\frac{1}{\chi_{\rm Lin} (\eta)},
%\right].
\end{equation} 
where $\eta=k/\left(2k_F^0\right)$ is a wavenumber in the units of the Fermi wavenumber $k_F^0=(3\pi^2 n_0)^{1/3}$ (with $n_0$ being the mean electron density) and $\mathfrak{F}[...]$ denotes the Fourier transformation operator.
For example, so-called semilocal KE functionals can be build by using the long wavelength expansion of Eq. (\ref{eq:start}) at $\eta<1$ \cite{ke_test_jctc}.

The engineering of advanced KE functionals is accompanied by involved theoretical manipulations with the constrain that the Lindhard function is reproduced if one first takes the UEG limit  $n\to n_0$ and then  performs the Fourier transform according to the relation (\ref{eq:start}). For example, using a series expansion of $\chi_{\rm Lin}^{-1} (\eta)$ for $\eta <1$, one recovers a constant term (associated to the Thomas-Fermi functional) and corrections which come with powers of $\eta=\frac{kn_0}{2k_Fn_0}$ and $\eta^2=\frac{k^2n_0}{4k_F^2n_0}$ whose real-space counterparts are $ s(\vec r)={|\nabla n(\vec r)|}/\left(2k_F(\vec r) n (\vec r)\right)$ and $ q(\vec r)={|\nabla^2 n(\vec r)|}/\left(4k_F^2(\vec r)n (\vec r)\right)$ (where $k_F(\vec r)=(3\pi^2 n(\vec r)^{1/3}$ is the local Fermi wavenumber). Another example of the operations connecting a real space density inhomogeneity with the density perturbation wavenumber in $\chi_{\rm Lin}^{-1} (\eta)$ is the functional integration using the local density  to map system properties locally on the UEG at the corresponding density \cite{LMGP,huang2010nonlocal}.
These schemes for the engineering of KE functionals based on Eq.~(\ref{eq:start}) can be symbolically  expressed as
\begin{equation}\label{eq:implies}
    \chi_{\rm Lin}(q;[n_0])\implies T_s[n(\vec r)].
\end{equation}

Being based on assumptions that can be theoretically justified to a certain degree, the KE functionals constructed on top of the Lindhard function  might not be traceable back numerically to Eq. (\ref{eq:start}) in the limit of the UEG leading to an inconsistency:
\begin{equation}\label{eq:n_implies}
    T_s[ n(\vec r)] \centernot\implies \chi_{\rm Lin}(q;[n_0]).
\end{equation}

Additionally, nonlocal KE functionals derived from Eq. (\ref{eq:start}) will  suffer from bias towards describing uniform densities best. It is yet unclear whether these functionals are capable of producing accurate density response for the UEG and realistic materials in the linear and beyond-linear response regimes.

In this work, we test some of the open questions mentioned above by implementing a direct perturbation approach. The electronic system is harmonically perturbed allowing us to compute the static density response function of UEG and any bulk material using OF-DFT as well as KS-DFT. This direct perturbation approach constitutes an effective tool to check inconsistencies in the sense of Eq. (\ref{eq:n_implies}) and generally the quality of the density response in the linear and beyond-linear regimes. We specifically focus on the case of fully degenerate electrons with a temperature  $T\ll T_F$, where $T_F$ is the Fermi temperature.   

A particular focus of this work is on KE functionals designed imposing Eq. (\ref{eq:start}). Since the Lindhard function based construction is not optimal for bound states and systems with band gaps, other strategies such as utilizing  the jellium-with-gap model \cite{PhysRevB.95.115153}, constrains on the Pauli potential and constraints on atom densities (e.g. Kato cusp condition) have been used.
These observations motivate us to employ the method of the direct perturbation approach to compute the static density (linear and beyond-linear) response function for various KE functionals to inspect the severity of the bias introduced by Eq. (\ref{eq:start}).
We reiterate that, in addition to the non-interacting kinetic energy (KE) functional, the XC functional needs to be supplied as an input of any DFT simulation. However, because we focus here on KE functionals, we purposely do not test the variability of our results with respect to different choices of XC functionals.

% whether the constraint (\ref{eq:start}) is respected in the required range of wavenumbers and  

% For example, let us consider  a weak harmonic perturbation of a free electron gas at a given wavenumber $\vec k$ in the linear response regime, $V_{\rm ext}\sim \cos(\vec k\cdot \vec r)$,  which generates a cosine-shaped density perturbation $n(\vec r)\sim \cos(\vec k\cdot \vec r)$. 
% In this case, it is clear that $\eta \neq s(\vec r)$ and $\eta^2 \neq q (\vec r)$, as it is illustrated in Fig. \ref{fig:wave}. 
% This simple example demonstrates the need for independent tools to crosscheck that the theoretical reasoning used to construct the KE functional do not violate the relation (\ref{eq:start}).

A particularly interesting problem to be analyzed using the direct perturbation method is related to the applicability of the UEG  based KE functional when the density perturbation is beyond the linear response regime. Indeed, when the perturbation is strong enough, linear density response theory (like $\chi_{\rm Lin}$) can be grossly inaccurate for the description of the density perturbation \cite{Dornheim_PRL_2020,Dornheim_PRR_2021, Moldabekov_JCTC_2022,Bohme_PRL_2022}.
Here we show that one can analyze the performance of the KE functionals by observing of the response from approximate OF-DFT KE functionals against KS-DFT results with increasing amplitude of the harmonic perturbation.
%Here  KS-DFT simulation results  can serve as a reference data to gauge the quality of the OF-DFT results.
%More specifically, we show that one can isolate errors in the density due to approximations in the KE functional by performing a comparative analysis with respect to KS-DFT data with  the XC functional being set to zero.

We consider an array of KE functionals, for example the nonlocal WT functional \cite{WT_paper} and MGP functionals \cite{MGP} (the latter uses functional integration techniques to define the kernel of the nonlocal term), the Laplacian-level, meta-GGA PGSL functional \cite{PGSL}, and LWT and LMGP functionals created by introducing a local density dependence into the kernels of MGP and WT, respectively \cite{LMGP}.
The constraint (\ref{eq:start}) was used  for the construction of these functionals. By design, the WT and LWT (and to an extent also MGP and LMGP) should satisfy Eq. (\ref{eq:start}), in principle, for all wavenumbers and PGSL should satisfy Eq. (\ref{eq:start}) at $\eta\lesssim 0.5$ (i.e., $k\lesssim k_F^0$). %We use that the direct perturbation approach  to test these properties. 
%In fact 
%In practice, we observe that some of the considered KE functionals violate  the constraint (\ref{eq:start}) at relevant wavenumbers. Additionally, we present an analysis of the behavior of the considered  KE functionals as the density perturbation increases beyond the linear-response regime such that $\chi_{\rm Lin}(\eta)$ becomes inaccurate for the description of the density perturbation.  

The application of the direct perturbation approach to real materials is provided for several phases of bulk Aluminum (Al) and Silicon (Si). We analyzed the density response of the valence electrons to a weak external harmonic field at different wavenumbers. This allowed us to relate the performance of the KE functionals in the UEG limit to the quality of the density response description for these realistic, inhomogeneous materials.

Finally, taking into account the interest in the application of OF-DFT for warm dense matter research \cite{Diaw2017, Cross_2014, Moldabekov_SciPost_2022, zhandos_pop18, Graziani_CPP_2022, pop_qhd_bonitz, white2013orbital}, we formulate recommendations for the application of these functionals at such extreme densities and temperatures for degenerate electrons.

The paper is organized as follows: In Sec. \ref{s:theory}, we provide a brief theoretical background of the considered KE functionals to emphasise the relevant physical aspects. We also introduce the direct perturbation method.  We provide the simulation details in  Sec. \ref{s:smd}. The results and discussions are presented in Sec. \ref{s:all_results} where we present static (linear and beyond-linear) response simulations for the UEG and bulk Al and Si.
The paper is concluded by summarizing main findings and providing an outlook over potential future works.

%%%%%%%%%%%%%%%%%%%%%%%%%%%%%%%%%%%%%%%%%%%%%%%%%%%%%%%%%%%%%%%%%%%%%%%%%%%%%%%%
% Theory
%%%%%%%%%%%%%%%%%%%%%%%%%%%%%%%%%%%%%%%%%%%%%%%%%%%%%%%%%%%%%%%%%%%%%%%%%%%%%%%%

\section{Theoretical background}\label{s:theory}

In this section, we first briefly introduce the nonlocal and Laplacian-level KE functionals considered in this work. Then, we discuss how the constraint (\ref{eq:start}) is used in their construction. Lastly, we introduce the harmonic perturbation approach.

\subsection{Nonlocal and Laplacian-level KE Functionals}\label{ss:KE}
The application of the constraint (\ref{eq:start}) is clearly illustrated by using the WT ansatz for the non-interacting KE density functional,
\begin{equation}\label{eq:KE_WT}
\begin{split}
   T_{\rm WT}[n(\vec r)]&=T_{\rm TF}[n(\vec r)]+T_{vW}[n(\vec r)]\\
   &+\int \int {\mathrm{d}}\vec r {\mathrm{d}}\vec r^{\prime}~ n(\vec r)^{5/6} K(\vec r-\vec r^{\prime};n_0)n(\vec r^{\prime})^{5/6},
   \end{split}
\end{equation}
where  $T_{\rm TF}[n(\vec r)]=\int {\mathrm{d}}\vec r~ \tau_{\rm TF}[n(\vec r)]$ is the ground state Thomas-Fermi KE with the kinetic energy density $\tau_{\rm TF}[n(\vec r)]=(3/10) (3\pi^2)^{2/3}n^{5/3}(\vec r)$ and $T_{vW}[n(\vec r)]=\int {\mathrm{d}}\vec r~  \left|\nabla n(\vec r) \right|^2/\left(8n(\vec r)\right)$ is the  von Weizs\"acker (vW) gradient correction, and the KE kernel $K(\vec r-{\vec r}^{\prime};n_0)$ is obtained in Fourier space by using Eq. (\ref{eq:start}),

\begin{equation}\label{eq:K_WT}
% \begin{split}
    \widetilde K(k;n_0)=\left[-\chi_{\rm Lin}^{-1}(k)+\chi_{\rm TF}^{-1}(k)+\chi_{\rm vW}^{-1}(k) \right] \frac{18}{25} n_0^{-1/3}\, .
% \end{split}
\end{equation}

Here $\chi_{\rm TF}^{-1}(k)=-\pi^2/(3\pi^2 n_0)^{1/3}$ is the Tomas-Fermi response function and and  $\chi_{\rm vW}^{-1}(k)=-k^2/(4n_0)$ is the vW contribution (e.g., see Refs. \cite{zhandos_pop18, Sjostrom_PRB2013}). The variational minimization of the WT KE functional  for the UEG under the constraint of a constant
particle number automatically reproduces $\chi_{\rm Lin}^{-1}(k)$ for all wavenumbers. Therefore, the WT KE functional is described as nonlocal.  We note that, in quantum plasma applications, the potential generated by the $T_{vW}[n(\vec r)]$ term in Eq. (\ref{eq:K_WT}) is commonly referred to as the Bohm potential \cite{zhandos_pop18, Manfredi_2008_prb}. 

Other nonlocal KE functionals based on the Lindhard function and considered in this work are MGP \cite{MGP}, LMGP and LWT \cite{LMGP}.
In contrast to the WT model  for which the KE potential $\delta T_{\rm WT}/\delta n$ is computed using an ansatz (\ref{eq:KE_WT}) for the KE functional,
the MGP potential is calculated by functional integration \cite{MGP}. 
The difference between MGP and LMGP is the way how the Lindhard function is used.
In the MGP, $\chi_{\rm Lin}^{-1}(k)$ is computed using the mean density $n_0$ of the valence electrons.
In the case of the LMGP, $\chi_{\rm Lin}^{-1}(k)$ is computed for each grid point using the density value $n_0\to n(\vec r)$ on this grid point.
Similarly, the LWT potential is computed using an $n_0\to n(\vec r)$ mapping locally for the WT KE kernel (\ref{eq:KE_WT}) \cite{LMGP}.
Therefore, being nonlocal functionals based on the Lindhard function and with rather non-trivial mathematical manipulations, the  MGP, LMGP, and LWT functionals are perfect candidates to showcase the utility of the direct perturbation approach for computing the density response and for testing functionals (i.e. in the limit of the UEG). 

In addition to the aforementioned nonlocal KE functionals, we consider three semilocal KE functionals.
We use a Laplacian-meta-GGA level KE functional, PGSL, developed by Constantin \textit{et al.}~\cite{PGSL} (with PGSL standing for ``Pauli-Gaussian second order and Laplacian'').
The PGSL KE functional used in this work has the following form:

\begin{equation}\label{eq:KE_PGSL}
\begin{split}
   T_{\rm PGSL}[n(\vec r)]&=T_{vW}[n(\vec r)]+\int {\mathrm{d}}\vec r ~\tau_{\rm TF}[n(\vec r)] F(s(\vec r), q(\vec r)),
   \end{split}
\end{equation}
where
\begin{equation}\label{eq:F_PGSL}
% \begin{split}
    F_{\rm PGSL}(s(\vec r), q(\vec r))=\exp(-\mu s^2(\vec r))+\beta q^2(\vec r),
   % \end{split}
\end{equation}
with $\mu=40/27$ and $\beta=0.25$.

The PGSL KE functional (\ref{eq:F_PGSL}), by its design, should reproduce the Lindhard density response function of the UEG at $k<2k_F^0$. In Ref. \cite{PGSL}, it was demonstrated that the PGSL KE functional provides accurate results for the bulk  properties of metals and semiconductors without using system-dependent parameters.

The second  semilocal KE functional that we use is PGS (standing for ``Pauli-Gaussian second order'') \cite{PGSL}, which has the same functional form as Eq. (\ref{eq:KE_PGSL}) but with the following kernel:
\begin{equation}\label{eq:F_PGS}
% \begin{split}
    F_{\rm PGS}(s(\vec r), q(\vec r))=\exp(-\mu s^2(\vec r)).
   % \end{split}
\end{equation}

The PGS functional is designed to satisfy the second-order gradient expansion following from the long wave length expansion of the inverse Lindhard function \cite{Perrot, zhandos_pop18, Moldabekov_pop2015, zmcpp2017}. 

The third  semilocal KE functional that we consider is the generalized gradient approximation by Luo, Karasiev, and Trickey (LKT) \cite{ground_LKT}:
\begin{equation}\label{eq:KE_PGSL}
\begin{split}
   T_{\rm LKT}[n(\vec r)]&=\int {\mathrm{d}}\vec r ~\tau_{\rm TF}[n(\vec r)] F_{\rm LKT}(s(\vec r), q(\vec r)),
   \end{split}
\end{equation}
where
\begin{equation}\label{eq:F_PGSL}
% \begin{split}
    F_{\rm LKT}(s(\vec r), q(\vec r))=\frac{1}{\cosh\left(as(\vec r)\right)}+\frac{5}{2}s^2(\vec r),
   % \end{split}
\end{equation}
with $a=1.3$.

The LKT functional is designed to satisfy the positivity of the Pauli functional and potential,  and was tested successfully on simple metals and semiconductors \cite{ground_LKT}.

\subsection{Harmonic Perturbations}\label{ss:direct_per}

The OF-DFT calculations of the electron density distribution  are performed by minimisation of the total energy under the constraint of a constant
particle number. The %used Lagrangian of free electron gas reads
corresponding Lagrangian is given by
\begin{equation}\label{eq:L}
\begin{split}
    L[n]=&T_{\rm s}[n]+W_{\rm H} [n]+W_{\rm ei} [n] +V_{\rm xc}[n]\\
    &+\int \mathrm{d}\vec r~2A\cos(\vec k\cdot\vec r)n(\vec r)-\mu_c\left(\int \mathrm{d}\vec r~ n(\vec r) - N \right),
\end{split}
\end{equation}
where $W_{\rm H} [n]$ is the classical Coulomb repulsion between electrons in a mean-field (Hartree) approximation, $W_{\rm ei} [n]=\int v_{\rm PP}(\vec r) n(\vec r) d\vec r$ is the potential energy due to the electron-ion interaction ($v_{\rm PP}$ is an ionic pseudo-potential), $V_{\rm xc}[n]$ is the XC energy, 
and $\mu_c$ is a constant defining the chemical potential at a fixed number of electrons $N$ in the simulation cell. Additionally, the Lagrangian (\ref{eq:L}) has a term corresponding to the external harmonic
perturbation with an amplitude $A$ and wavenumber $\vec k$.

The latter leads to a deviation of the electron density from its mean value,
\begin{equation}
 \delta n_{A,\vec k}(\vec r)=n_{A,\vec k}(\vec r)-n_0.   
\end{equation}

For the UEG, in the linear response regime (i.e., for weak perturbations), the change in the density $\delta n_{A,\vec k}(\vec r)$ can be computed using the static electronic density response function $\chi(\vec k)$,

\begin{equation}\label{eq:chi}
    \delta n_{A,\vec k}(\vec r)=\chi(\vec k) 2A\cos(\vec k\cdot\vec r).
\end{equation}

Therefore, the calculation of the electron density perturbation  $\delta n_{A,\vec k}(\vec r)$ due to an external harmonic field and the inversion of Eq. (\ref{eq:chi})  allow one to compute the static density response function $\chi(\vec k)$. We note that this method can be generalized to inhomogeneous systems  and to the dynamic density response function \cite{Moldabekov_dft_kernel, moldabekov2023linearresponse, moldabekov2023averaging}.

Next, if we neglect the XC energy and set $V_{\rm xc}=0$ in Eq. (\ref{eq:L}), the OF-DFT calculation of the harmonically perturbed system delivers the screened non-interacting density response function,
with the screening effect being due to the  presence of the Hartree term in the Lagrangian (\ref{eq:L}).
For this case,  the exact analytical solution in thermodynamic limit reads~\cite{quantum_theory}:
\begin{equation}\label{eq:rpa}
    \chi_{\rm RPA}( k)=\left[1-\chi_{\rm s}( k)\frac{4\pi}{k^2}\right]^{-1}\chi_{\rm s}( k),
\end{equation}
where $\chi_s$ is the static Kohn-Sham response which reduces to $\chi_{\rm Lind}$ for the UEG and the inverse reduces to a simple ``one-over'' operation. The resulting response in Eq. (\ref{eq:rpa}) is referred to as random phase approximation (RPA).

In this way, using the OF-DFT results for the density response function of the UEG in the RPA and Eq. (\ref{eq:rpa}), we can directly test whether a given KE functional is able to reproduce the Lindhard function at the relevant wavenumbers.

Increasing the amplitude of the external perturbation $A$ eventually leads to a density perturbation outside of the linear response domain. We recall that the KS-DFT method delivers an exact result for a given choice of XC functional (including omitting it completely to obtain the so-called RPA approximation) .
Therefore, by performing a comparative analysis of the OF-DFT and KS-DFT results for the density perturbation beyond the linear response regime, we can rigorously assess to which degree the quality of the KE functionals built on the basis of the linear response function of the UEG deteriorates. In general,  OF-DFT is used to describe inhomogeneous systems. Therefore, we will also provide examples for its application to a system that contains ions (bulk Al and Si). %The direct perturbation approach allows one to investigate the performance of KE functionals in detail by varying the amplitudes and wavenumbers of the external harmonic perturbation.
For context, in prior works, the direct perturbation approach has been used to compute the static density response function and XC kernel of  warm dense matter using quantum Monte-Carlo \cite{Bohme_PRL_2022, Dornheim_PRL_2020} and thermal KS-DFT methods \cite{Moldabekov_dft_kernel, Moldabekov_JCTC_2022, hybrid_results, Moldabekov_JCP_2021, Moldabekov_PRB_2022}. Here, we extend its use to probe the quality of KE density functionals in OF-DFT.

We note that the response functions considered so far are frequency independent, that is, they describe the density response due to an {\it adiabatic} perturbation whereby the electrons have an infinite time to adjust to the applied perturbation. The ability of the KE functional approximations to produce quality adiabatic responses is key to quality time-dependent OF-DFT simulations \cite{jiang2022efficient,covington2021coupled,della2022orbital}. In Ref. \citenum{jiang2022efficient} Jiang {\it et al.} argue that, while nonlocal functionals (specifically LMGP) provide good approximations to the adiabatic response, semilocal, GGA functionals also deliver an accurate adiabatic response. The test systems in that work were comprised of clusters of various sizes. By considering systems with periodicity, this work aims at providing valuable information as to whether (semi)local KE functionals live up to the expectations set for in Ref. \citenum{jiang2022efficient}.

\section{Calculation parameters}\label{s:smd}

The OF-DFT simulations were performed using the DFTpy code that is based on a plane-wave expansion of the electron density \cite{DFTpy}.
The KS-DFT calculations were performed using the ABINIT package \cite{Gonze2020, Romero2020, Gonze2016, Gonze2009, Gonze2005, Gonze2002}.
We simulated a free electron gas in the ground state  at a characteristic metallic density $n_0= 2\times 10^{23}~{\rm cm^{-3}}$ with periodic boundary conditions.

\begin{figure*}\centering
\includegraphics[width=0.8\textwidth]{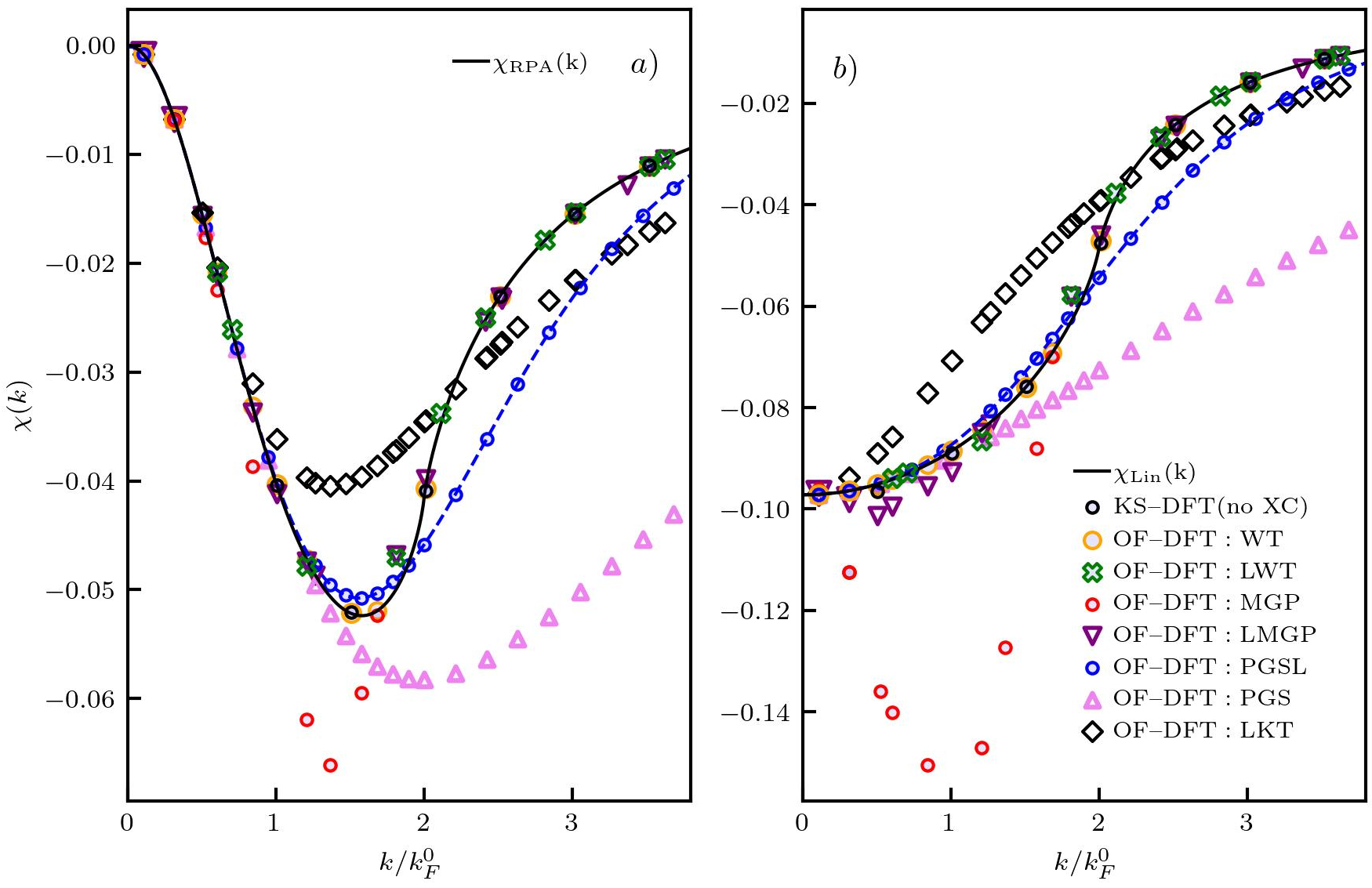}
\caption{\label{fig:chi_nullxc} Linear density response function of the UEG in the ground state at metallic density a) with screening (RPA) and b) without screening.  The results are computed using OF-DFT with different KE functionals and KS-DFT setting the XC potential to zero. The solid line represents the exact analytical results. The simulation data are computed for $N=7168 $, $N=66$, $N=38$ and $N=14$ electrons in the simulation cell.
}
\end{figure*} 

For the calculation of the linear response function, the amplitude of the perturbation is set to $A=0.01$ (in Hartree), which creates a weak density perturbation described accurately by linear response theory.
To test the KE functionals beyond the linear response regime, we consider $A=0.1$, $A=0.5$, and $A=1$.
The wavenumber of the external perturbation is set along the $x$-axis. The wavelength of the perturbation has to be commensurate with the size $L$ of the main simulation cell, which is defined by the relation  $NL^3=n_0$. Accordingly, the wavenumber values of the external harmonic perturbation are given by  $k=j\times k_{\rm min}$, where $k_{\rm min}=2\pi/L$ and $j$ is a positive integer number.
We reconstruct the density response function dependence on $k$ in the range $0.1\leq k/k_F^0<4$ by performing calculations for different $j$ values and for different numbers of particles.

The OF-DFT calculations of free electron gas were performed for $7168$, $66$, $38$, and $14$ electrons in the simulation cell. 
For $7168$ particles we have $k_{\rm min}/k_F^0\simeq 0.105$. The grid spacing was set to $L/200$.
% , corresponding to a kinetic energy cutoff of $600~{\rm eV}$.

The following KE density functional approximants were considered: WT \cite{WT_paper},  MGP \cite{MGP}, LWT and LMGP \cite{LMGP}, PGSL \cite{PGSL}, PGS \cite{PGSL}, and LKT \cite{ground_LKT}. Furthermore, we consider two cases: XC functional being set to LDA  \cite{Perdew_Wang_PRB_1992} and the case where the XC functional is omitted (RPA).

% The total energies were converged  to within $10^{-6}\times N~\rm eV ?$.

The  KS-DFT simulations were performed  for $N=38$ electrons in  the main cell with $40$ bands and with $30~{\rm Ha}$ energy cutoff. 
The $k$-point sampling was set to $10\times10\times10$. We present the results from KS-DFT simulations with zero XC functional and with LDA  \cite{Perdew_Wang_PRB_1992}.
We cross-checked the KS-DFT  results by reproducing our ABINIT data with independent calculations based on the GPAW code~\cite{GPAW1, GPAW2, ase-paper, ase-paper2}.

For the simulation of the valence electrons in a bulk of Al and Si, we used the unit primitive cells of the corresponding lattice structures. The presented KS-DFT simulation results are obtained using Quantum ESPRESSO (QE) \cite{Giannozzi_2009, Giannozzi_2017} implemented in the Python engine QEpy \cite{qepy}. The  energy cutoff was set to  $30~{\rm Ha}$ (corresponding to a wavefunction plane wave cut-off energy of 816 eV) and the $k$-point sampling was set to $16\times16\times16$. The QE results for a considered setup are cross checked by the GPAW simulations. OF-DFT simulations for Al using DFTpy were run with a kinetic energy cutoff of $240~{\rm eV}$.  Both KS-DFT and OF-DFT calculations for Al were run with BLPS local pseudopotentials \cite{huang2008transferable}.

\section{Results}\label{s:all_results}

We first present results for the static linear response function of the electron gas computed using an external harmonic field with amplitude $A=0.01$ at wavenumbers $0.1\leq k/k_F^0<4$.
Then, we present an analysis of the accuracy of OF-DFT calculations for the strongly perturbed (beyond linear response) electron gas.

\subsection{Linear density response} \label{ss:results_A0_01}

Let us first consider the ideal electron gas neglecting all XC effects. 
In Fig. \ref{fig:chi_nullxc}, we present the results for the density response function computed using different KE functionals and setting the XC functional to zero. In Fig. \ref{fig:chi_nullxc}a), we compare the OF-DFT results with the screened (RPA) Lindhard density response function (\ref{eq:rpa}).
Additionally, in Fig. \ref{fig:chi_nullxc}a), we show the data computed using KS-DFT with zero XC functional, but non-zero Hartree mean-field potential.
We observe that, as expected, the KS-DFT data for ideal electron gas accurately reproduce the analytical RPA result. The same is the case for the OF-DFT data based on the WT KE functional.
In contrast, we see that the MGP KE functional based OF-DFT data significantly deviate from the exact RPA data at $1\lesssim k/k_F^0 \lesssim 2$.
The MGP based results show substantial disagreements with the exact RPA result at $0.5\lesssim k/k_F^0 \lesssim 2$. The design choices leading to the MGP kernel are such that the kernel should be optimized (i.e., optimal values for two parameters defining the so-called ``kinetic electron''~\cite{MGP}). The results presented here show that without optimization MGP fails to deliver correct response across a wide array of wavevectors. Fig. \ref{fig:chi_nullxc} shows that the LMGP results are in good agreement with the exact RPA data at all considered wavenumbers. This is in line with the design choices that led to the formulation of LMGP which were to yield a KE functional based on MGP with no adjustable parameters~\cite{LMGP}. 
%For the region corresponding to the minimum of the RPA density response function at $k\approx 1.5 k_{F}^0$, the LMGP results are not accessible due to a numerical instability leading to a constant $\chi(k)$.  
Among the nonlocal functionals, we note that the LWT functional (by design) is computed by using  the local density approximation in the KE kernel $n_0\to n(\vec r)$~\cite{LMGP}. The static density response function computed using 38 and 24 particles show close agreement of the LWT based data with WT based results%., where the functional integration range with respect to density is set to be between $0.1n_0$ and $2.8n_0$. 
In the case of other considered smaller and larger particle numbers, the LWT became numerically unstable at intermediate wavenumbers ($0.5\lesssim k/k_F^0 \lesssim 2$)  showing large deviations from the WT based data for the UEG.  To avoid possible spurious results, we will not consider LWT for the density response in more non-trivial cases of strong perturbations and real materials. 

The PGSL and PGS functional based results are in close agreement with the exact RPA data at $k\lesssim 2k_F^0$ and at $k\lesssim k_F^0$.
This is in agreement with the analytically enforced constraints to the PGSL and PGS.
In general, all considered KE functionals in Fig. \ref{fig:chi_nullxc}a) closely agree with the exact solution (\ref{eq:rpa}) at $k<k_F^0$.
This can be understood by observing that the screening leads to  $\chi_{\rm RPA}(k) \approx -k^2/4\pi$  in Eq. (\ref{eq:rpa}) at small wavenumbers since $\chi_{\rm Lin}(k\to0)\to {\rm const}$~\cite{kugler_bounds}. Therefore, the Coulomb term $4\pi/k^2$ in the denominator of Eq. (\ref{eq:rpa})---representing screening effect---dominates the $k$ dependence of the $\chi_{\rm RPA}(k)$ at small wavenumbers. This can mask the inaccuracies of the OF-DFT results at $k<k_F^0$.

 \begin{figure}\centering
\includegraphics[width=0.4\textwidth]{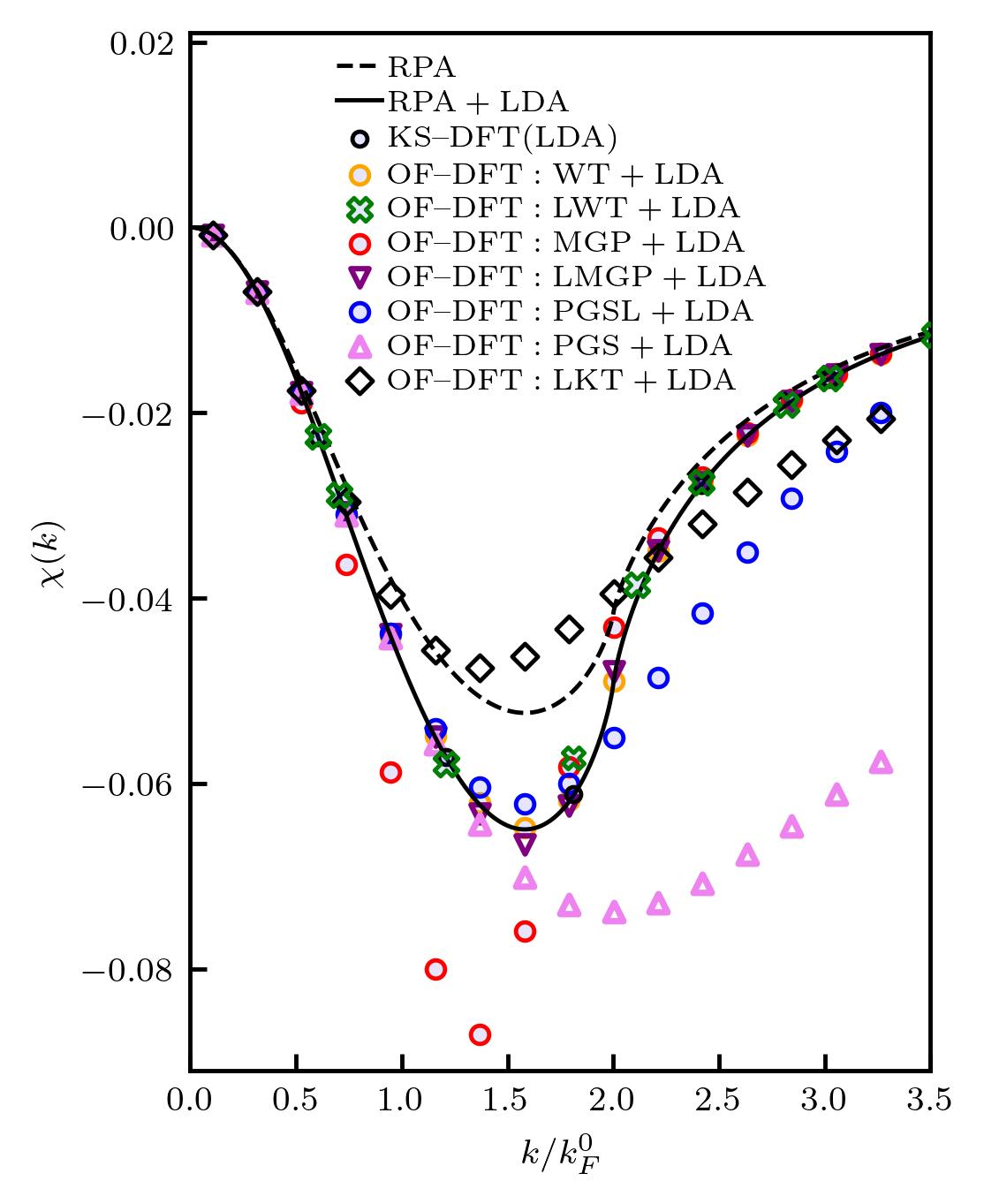}
\caption{\label{fig:chi_lda}  Density response function of the correlated UEG in the ground state at metallic density computed using OF-DFT with different KE functionals and KS-DFT with the LDA XC functional.
The dashed line represents the analytical result Eq. (\ref{eq:chi_lda}). 
The results are computed for $N=7168$ electrons in the simulation cell.
}
\end{figure}

\begin{figure}[t!]\centering
\includegraphics[width=0.48\textwidth]{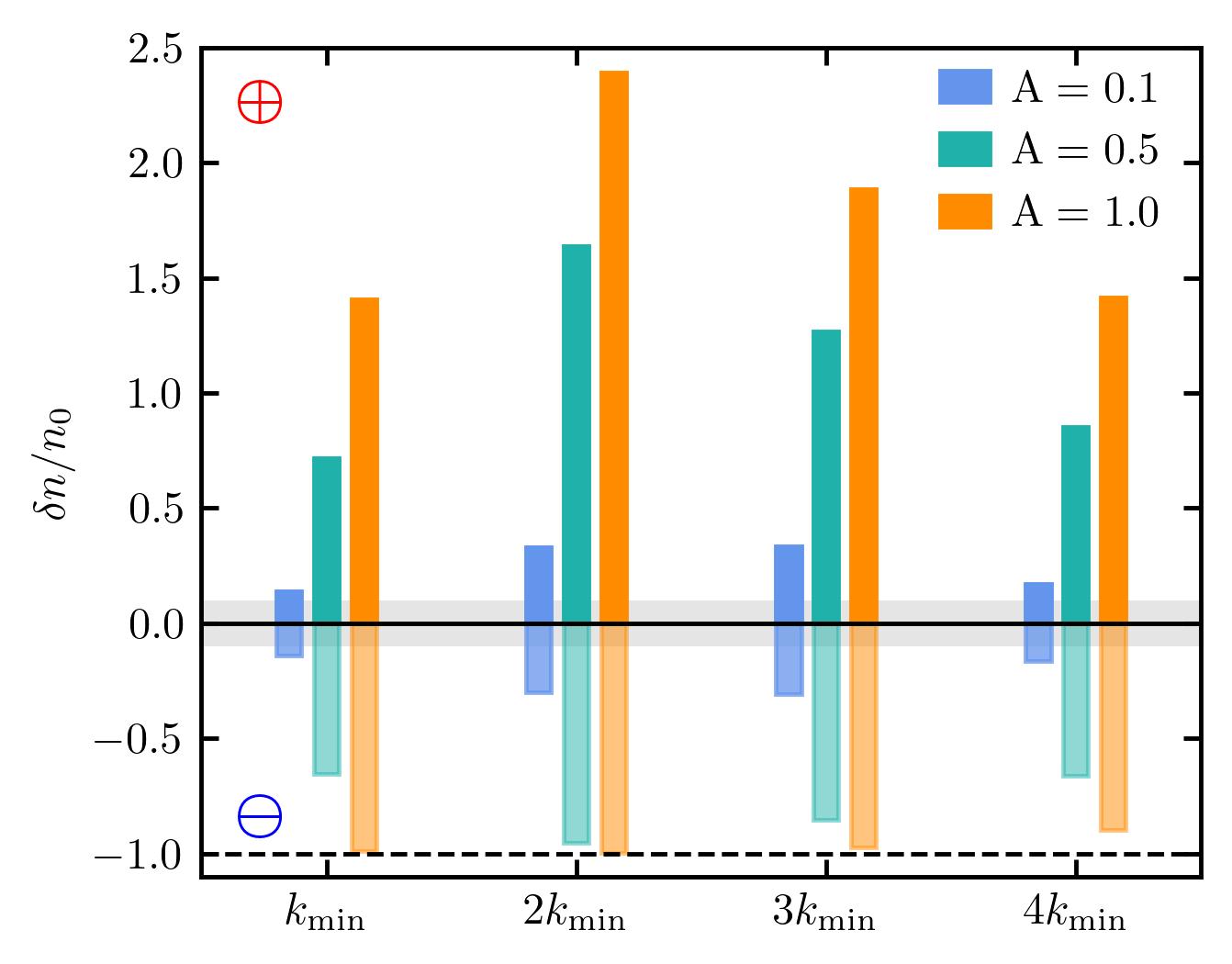}
\caption{\label{fig:per_den} 
Density perturbation magnitude at different external field amplitudes and wavenumbers in the region with $\delta n(\vec r)>0 $ denoted as $\oplus$ (positive values) and in the region with $\delta n(\vec r)<0$  denoted as $\ominus$ (negative values). The perturbation wavenumber is set to  $k=k_{\rm min}\simeq 0.604 k_F^0$, $k=2k_{\rm min}$, $k=3k_{\rm min}$, and $k=4k_{\rm min}$. The shaded area shows $|\delta n (\vec r)| < 0.1 n_0$ values.  The dashed horizontal line represents $n(\vec r)=0$.  The data is computed from KS-DFT simulations with zero XC functional for $N=38$ particles in the simulation cell.
}
\end{figure}

In order to assess the quality of the ideal density response function without screening,
we invert Eq. (\ref{eq:rpa}) with $\chi_{\rm RPA}(k)=\chi_{\rm RPA}^{\rm DFT}(k)$ being computed using the considered KE functionals in OF-DFT and using KS-DFT data. The results obtained in this way are presented in Fig. \ref{fig:chi_nullxc}b).
Additionally, in Fig. \ref{fig:chi_nullxc}b), we compare the DFT results with the Lindhard function which defines the constraint (\ref{eq:start}) used to construct the considered KE functionals. 
In the $k\to0$ limit, all considered KE functionals reproduce the corresponding limit of the Lindhard function as it is illustrated in Fig. \ref{fig:chi_nullxc}b) for the point at $k=0.1k_F^0$ computed using $N=7168 $ particles. In fact, it is the limit of the TF model.
At large wavenumbers $k>2.5k_F^0$ dominated by the vW KE term, the LWT, MGP, and LMGP KE functionals are able to correctly describe the  density response function of the ideal electron gas. This is indeed the well-known single particle regime in which the vW functional provides the exact kinetic energy value. The WT KE functional accurately reproduces the Lindhard function at all considered $k$ values. 
% From Fig. \ref{fig:chi_nullxc}b), we see that the modifications introduced into the WT functional to construct the LWT functional violates the constraint (\ref{eq:start}) at $0.2 \lesssim k/k_F^0 \lesssim 2$.
Among other constrains, the MGP functional was designed to satisfy  the relation  (\ref{eq:start}) for, in principle, all wavenumbers. 
However, from Fig. \ref{fig:chi_nullxc}, we see that the MGP results, for the reasons mentioned before, do not satisfy  the relation  (\ref{eq:start})  at $0.1\lesssim k/k_F^0 \lesssim 2$.  In contrast and in line with the discussion above, the LMGP based ideal density response function is in good agreement with the Lindhard function at the considered wavenumbers. 
From Fig. \ref{fig:chi_nullxc}, and following the expected behavior, the PGSL functional reproduces the Lindhard function for $k\lesssim 2k_F^0$ and approaches the Lindhard function at large wavenumbers $k>3k_F^0$. In contrast, the PGS functional does not give the correct large wavenumber limit defined by the vW KE functional.
Finally, we find close agreement between the LWT and the WT based results. %  where density gradients tend to zero and $n(\vec r)\to n_0$ is realised with a high enough precision. 

Next, we consider an interacting electron gas using the LDA XC functional and the corresponding density response function $\chi_{\rm LDA}(k)$.
In Fig. \ref{fig:chi_lda}, we show the results for  $\chi_{\rm LDA}(k)$ computed using OF-DFT and KS-DFT for $N=38$ electrons in the main simulation cell.
Additionally, we compare the results with the exact solution for the UEG with the LDA XC functional,
\begin{equation}\label{eq:chi_lda}
    \chi_{\rm LDA}( k)=\frac{\chi_{\rm Lin}( k)}{1-\frac{4\pi}{k^2}\left[1-G_{\rm LDA}(k)\right]\chi_{\rm Lin}( k)},
\end{equation}
where $G_{\rm LDA}(k)\sim k^2$ is the local field correction of the UEG in the long wavelength limit.

To compute $G_{\rm LDA}(k)$,  we used the compressibility sum-rule~\cite{IIT_1987},
\begin{eqnarray}\label{eq:CSR}
G_{\rm LDA}(k)=\lim_{k\to0}G(k) = - \frac{k^2}{4\pi} \frac{\partial^2}{\partial n_0^2} \left( n_0 V_\textnormal{xc}[n_0] \right).
\end{eqnarray}
For completeness, we note that a numerical analysis of Eq.~(\ref{eq:CSR}) has been provided in the recent Ref.~\cite{Moldabekov_dft_kernel} both at ambient conditions and in the warm dense matter regime.

From  Fig. \ref{fig:chi_lda}, we observe that the WT based data is in perfect agreement with the exact solution (\ref{eq:chi_lda}).
The same is the case for the KS-DFT data, which validates the solution (\ref{eq:chi_lda}). 
The  MGP based results strongly deviate from the exact solution (\ref{eq:chi_lda}) and the KS-DFT data at $1\lesssim k/k_F^0 \lesssim 2$. LMGP results follow the benchmark result despite showing small deviation at around $1.5 \lesssim k/k_F^0 \lesssim 2$.
The PGSL (PGS) based OF-DFT data is in good agreement with the solution (\ref{eq:chi_lda}) and the KS-DFT results at $k/k_F^0 \lesssim 2$ ($k/k_F^0 \lesssim 1$).
These relative deviations of the OF-DFT results from the exact data take place at exactly same wavenumbers as for the ideal electron gas case shown in Fig. \ref{fig:chi_nullxc}a). 
Clearly, the  inaccuracies in the OF-DFT results originate from the approximations in the KE functional.
Therefore, we can isolate these errors from other possible effects by comparing the OF-DFT data and KS-DFT data computed with zero XC functional.
Although it is trivial to see for the UEG, in a strongly inhomogeneous case, various types of numerical error cancellations might mask to a certain degree the deficiencies of a KE functional. 

\begin{figure}\centering
\includegraphics[width=0.48\textwidth]{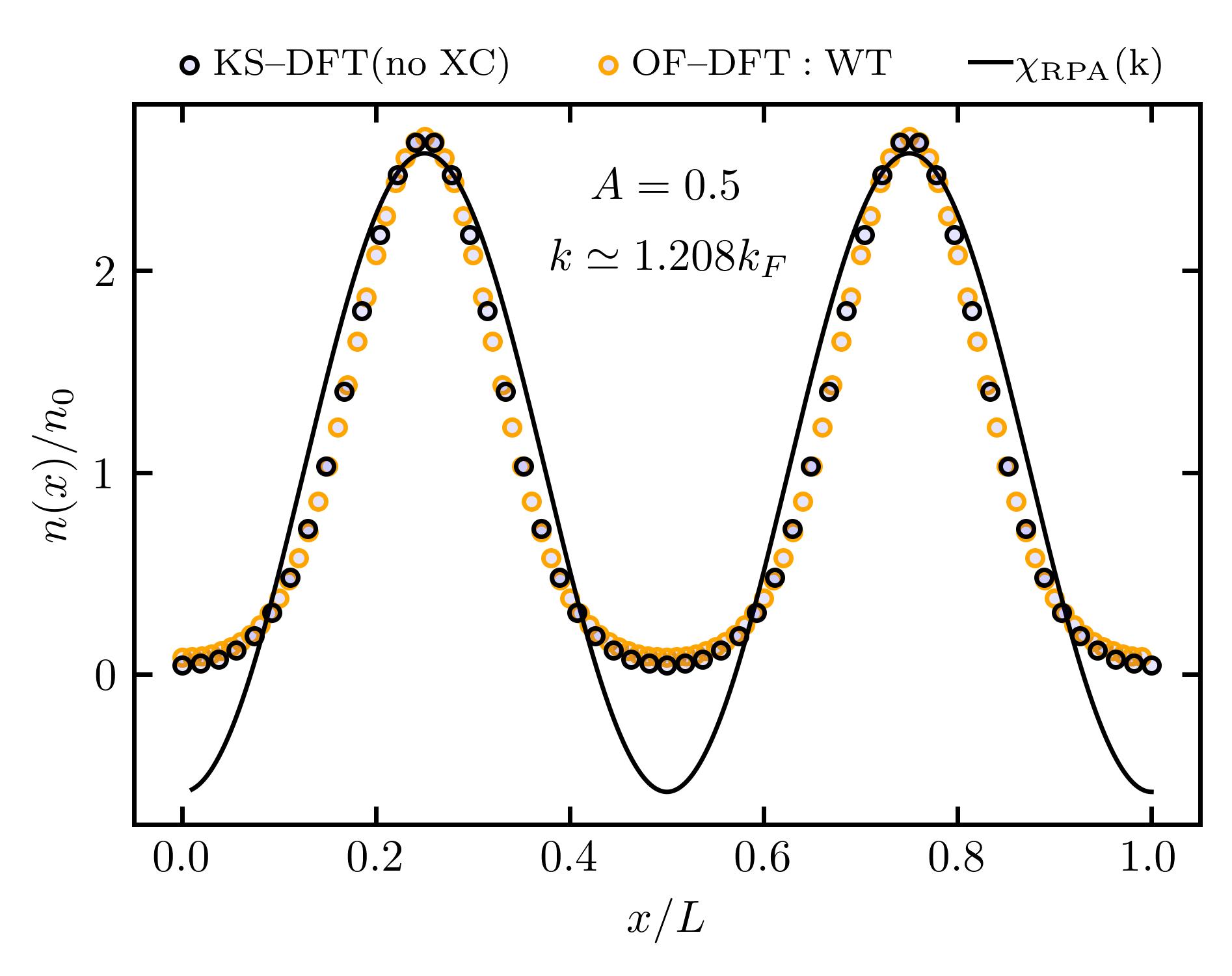}
\caption{\label{fig:den} Density profile along the perturbation direction at $A=0.5$ and $k\simeq 1.208 k_F^0$. The results are computed using OF-DFT with WT functional and KS-DFT and setting the XC functional to zero for both. The solid line is the density values $n=n_0+\delta n$ with $\delta n$ computed using the RPA density response function Eq. (\ref{eq:rpa}) of the UEG  in Eq. (\ref{eq:chi}).  
}
\end{figure}

  \begin{figure}\centering
\includegraphics[width=0.48\textwidth]{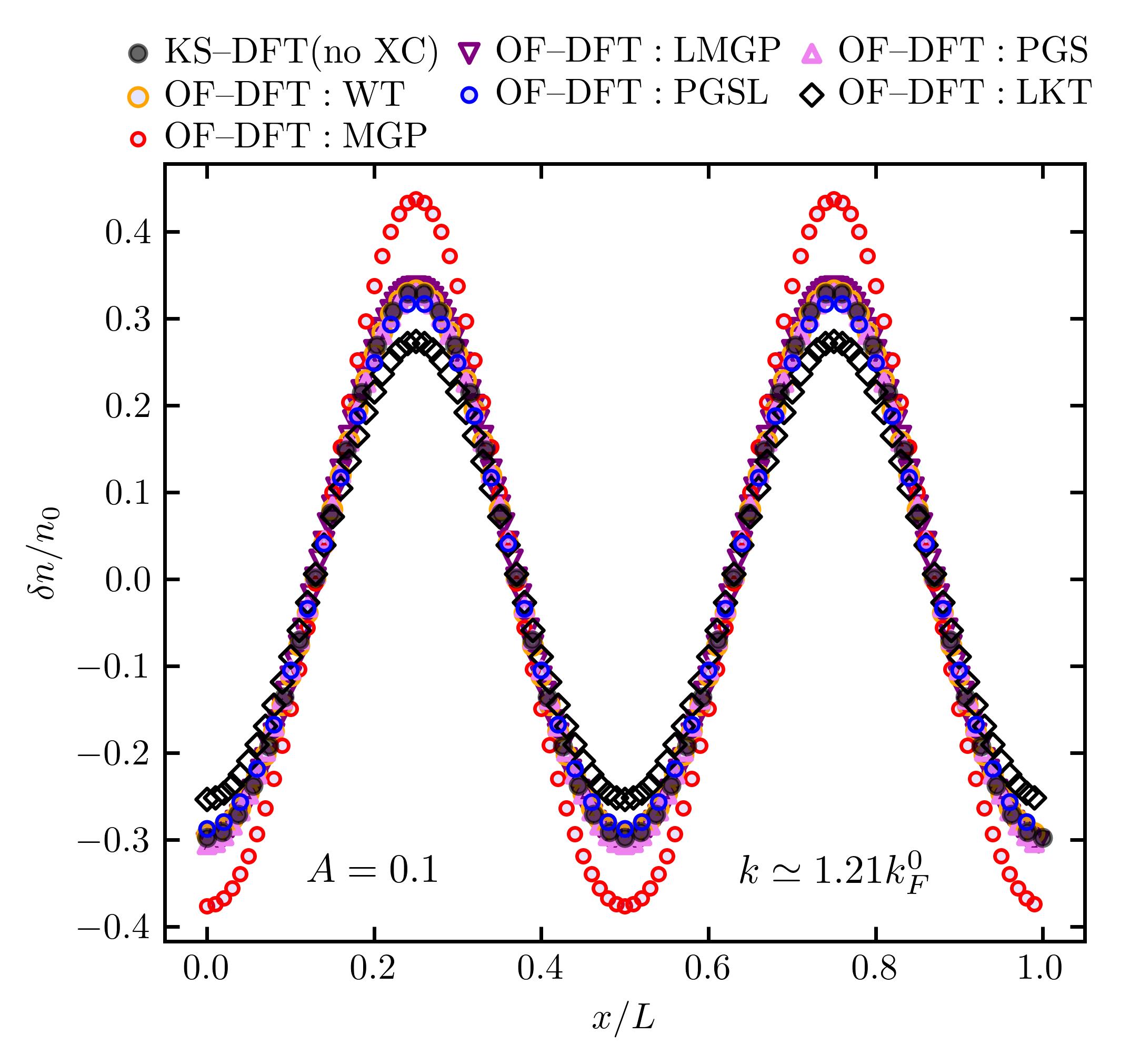}
\caption{\label{fig:den_2} Density perturbation profile $\delta n(\vec r)$  along the perturbation direction at $A=0.1$ and $k\simeq 1.208 k_F^0$. The results are computed using OF-DFT with different KE functionals and KS-DFT setting the XC potential to zero.
}
\end{figure} 

\begin{figure}\centering
\includegraphics[width=0.48\textwidth]{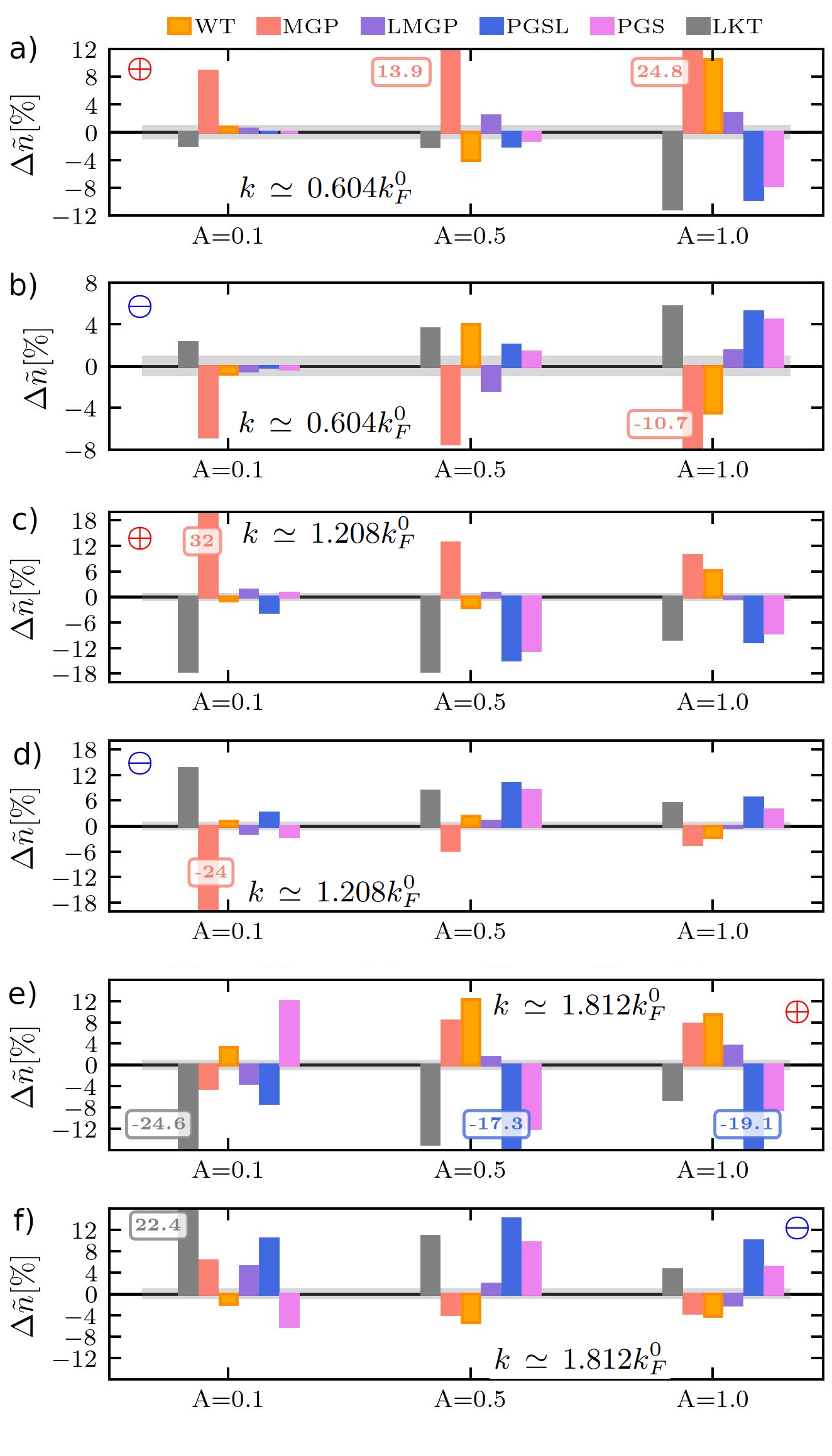}
\caption{\label{fig:bar_q1}  The largest value of the error in the relative density deviation of the OF-DFT data from the reference KS-DFT data, Eq. (\ref{eq:Dn}), in the region with $\delta n(\vec r)>0$ (top panel, denoted as $\oplus$) and in the  region with  $\delta n(\vec r)<0$ (bottom panel, denoted as $\ominus$) for $A=0.1$, $A=0.5$, and $A=1$ at $k\simeq 0.604 k_F^0$.
The grey area indicates $|\Delta \widetilde n_{\oplus}|\leq \pm 1~\%$ and $|\Delta \widetilde n_{\ominus}|\leq \pm 1~\%$.
The results are computed for the electron gas with XC potential set to zero.
}
\end{figure}

\subsection{Strongly perturbed electron gas}

Let us now consider a strongly perturbed electron gas. Specifically, we consider an electron gas with zero XC energy in order to analyze the errors due to the approximation of the KE functional in OF-DFT.
To generate strong deviations from the uniform case, we use $A=0.1$, $A=0.5$, and $A=1.0$.
In Fig. \ref{fig:per_den}, we show the values of the largest positive ($\delta n_{\oplus}$) and negative ($\delta n_{\ominus}$)  deviations of the density from the mean value $n_0$.
We note that the $|\delta n_{\ominus}|$ value is physically constrained to be smaller than $n_0$ since the electron density cannot have negative values.
In Fig. \ref{fig:per_den}, we present the results computed using the KS-DFT simulations with zero XC functional at wavenumbers of the external perturbation $k_{\rm min}$, $2k_{\rm min}$, $3k_{\rm min}$, and $4k_{\rm min}$.
The data is computed for $N=38$ electrons. Correspondingly, we have $k_{\rm min}=2\pi/L \simeq 0.604 k_F^0$.
The density perturbation values of less than $\pm0.1n_0$ are indicated by the grey area.
From Fig. \ref{fig:per_den}, one can see that, at $A=0.1$,  we have $|\delta n_{\oplus}|\geq 0.1 n_0$ and $|\delta n_{\ominus}|\geq 0.1 n_0$. 
These density deviations from the uniform case are already beyond the linear response regime \cite{Dornheim_PRR_2021,Dornheim_PRL_2020} and can be characterized as a strongly inhomogeneous electron gas.
A further increase in the amplitude of the external perturbation to $A=0.5$ leads to the magnitudes of the density perturbation  $|\delta n_{\oplus}|>n_0$ and $|\delta n_{\ominus}|> 0.5 n_0$.
At $A=1.0$, almost all electrons are localized in the density accumulation regions and  $|\delta n_{\ominus}|\approx n_0$ at $k_{\rm min}$, $2k_{\rm min}$, and $3k_{\rm min}$. 

Due to the use of the local density approximation and various constrains in KE functionals (such as the positivity of the density and  Pauli potential, and an exact single particle limit), OF-DFT is expected to describe strongly inhomogenous systems with a certain accuracy. This has motivated us to look at the density response of the electron gas to a strong external perturbation.
In Fig. \ref{fig:den_2}, we show the density profile  computed for the perturbation amplitude $A=0.5$ and the wavenumber $k=2k_{\rm min}$ using the WT functional in OF-DFT and using  KS-DFT with the XC functional set to zero in both. We compare the KS-DFT and OF-DFT results with the density values $n=n_0+\delta n$ from  the linear response theory with $\delta n$ being computed using the RPA density response function Eq. (\ref{eq:rpa}) of the UEG  in Eq. (\ref{eq:chi}).   From this figure, we see that the linear response theory gives nonphysical negative values for the density since $A=0.5$  is well beyond the weak perturbation regime~\cite{Dornheim_PRL_2020,POP_review}. In contrast, the OF-DFT results based on the WT functional are in close agreement with the KS-DFT results and, as expected, always positive. 
We do not consider higher order response functions of the UEG appearing in the Taylor expansion of the density perturbation with respect to weak external perturbation \cite{Mikhailov_PRL_2014, Tolias_EPL_2023}. In fact, the considered KE functionals are designed to reduce (analytically at all or certain wavenumbers) to the Lindhard function in the UEG limit. Nevertheless, it can be seen from a corresponding Taylor expansion that the first order density response dominates over contributions of higher order terms.  
As we show next, this correlates with the observation that the KE functionals reproducing correct first order density response (at a given wavenumber of the perturbation) provide more accurate results.

The profile of the density distribution along the perturbation direction is shown in Fig. \ref{fig:den} for $A=0.1$ and $k=2k_{\rm min}$ for both KS-DFT and OF-DFT simulations with different KE functionals.
We observe that the OF-DFT data computed using the WT, LMGP, PGSL, and PGS KE functionals are in good agreement with the KS-DFT data.
This clearly illustrates that the KE functionals based on the linear response function $\chi_{\rm Lin}(k)$ can remain valid beyond the weak density perturbation regime (given that they reproduce $\chi_{\rm Lin}(k)$  in the UEG limit).
In contrast, the MGP and LKT based results show significant errors in the density values compared to the KS-DFT data. This is not surprising since these functionals are not able to accurately describe the ideal density response function of the UEG at the considered wavenumber $k=2k_{\rm min}\simeq 1.21 k_F^0$ (see Fig. \ref{fig:chi_nullxc}a)).

To further analyze the quality of the considered KE functionals,
we present histograms of  the largest values of the error in the density accumulation and depletion regions at different values of the perturbation amplitude $A$ and wavenumber $k$ in Fig. \ref{fig:bar_q1}. We consider perturbation amplitudes $A=0.1$, $A=0.5$ and $A=1$ and wavenumbers  $k=k_{\rm min}\simeq 0.6k_F^0$ (panels a) and b)), $k=2k_{\rm min}$ (panels c) and d)), and $k=3k_{\rm min}$ (panels e) and f)).
To quantify the error in the electron density, we measure the relative density deviation of the OF-DFT results relative to the KS-DFT data,
\begin{equation}\label{eq:Dn}
    \Delta \widetilde n ~[\%]=\frac{\delta n_{\rm OF\text{-}DFT}-\delta n_{\rm KS\text{-}DFT}}{{\rm max}\left|\delta n_{\rm KS\text{-}DFT}\right|} \times 100.
\end{equation}

We compute the $\Delta \widetilde n$ values for the density accumulation ($\delta n(\vec r)>0$) and density depletion ($\delta n(\vec r)<0$) regions separately.
The former is presented in the panels labeled by $\oplus$ and the latter in the panels labeled by $\ominus$ in  Fig. \ref{fig:bar_q1}. The grey areas depict error values less than $\pm1\%$. From Fig. \ref{fig:bar_q1} we see that: The WT and LMGP KE functionals deliver rather accurate results with $\Delta \widetilde n<5\%$ for $\delta n/n_0 \leq 2.4$ and all considered wavenumbers; The PGSL and PGS KE functionals are accurate at $k\lesssim k_F^0$ for the harmonically perturbed inhomogeneous electron gas with  $\delta n/n_0\lesssim 0.5$; The LKT KE functional is accurate at $k\lesssim 0.5 k_F^0$ for the harmonically perturbed inhomogeneous electron gas with  $\delta n/n_0 \lesssim 0.5$; The MGP KE functional is accurate at $k\lesssim 0.5 k_F^0$ for the harmonically perturbed inhomogeneous electron gas with  $\delta n/n_0 < 0.5$ and cannot be considered as reliable at larger wavenumbers.

Summarizing the key findings from the data presented in  Fig. \ref{fig:bar_q1} and for the UEG limit, we conclude that the accuracy of a KE functional for the  inhomogeneous regime strongly correlates with its accuracy for the density response in the limit of  the UEG.

\begin{figure}\centering
\includegraphics[width=0.5\textwidth]{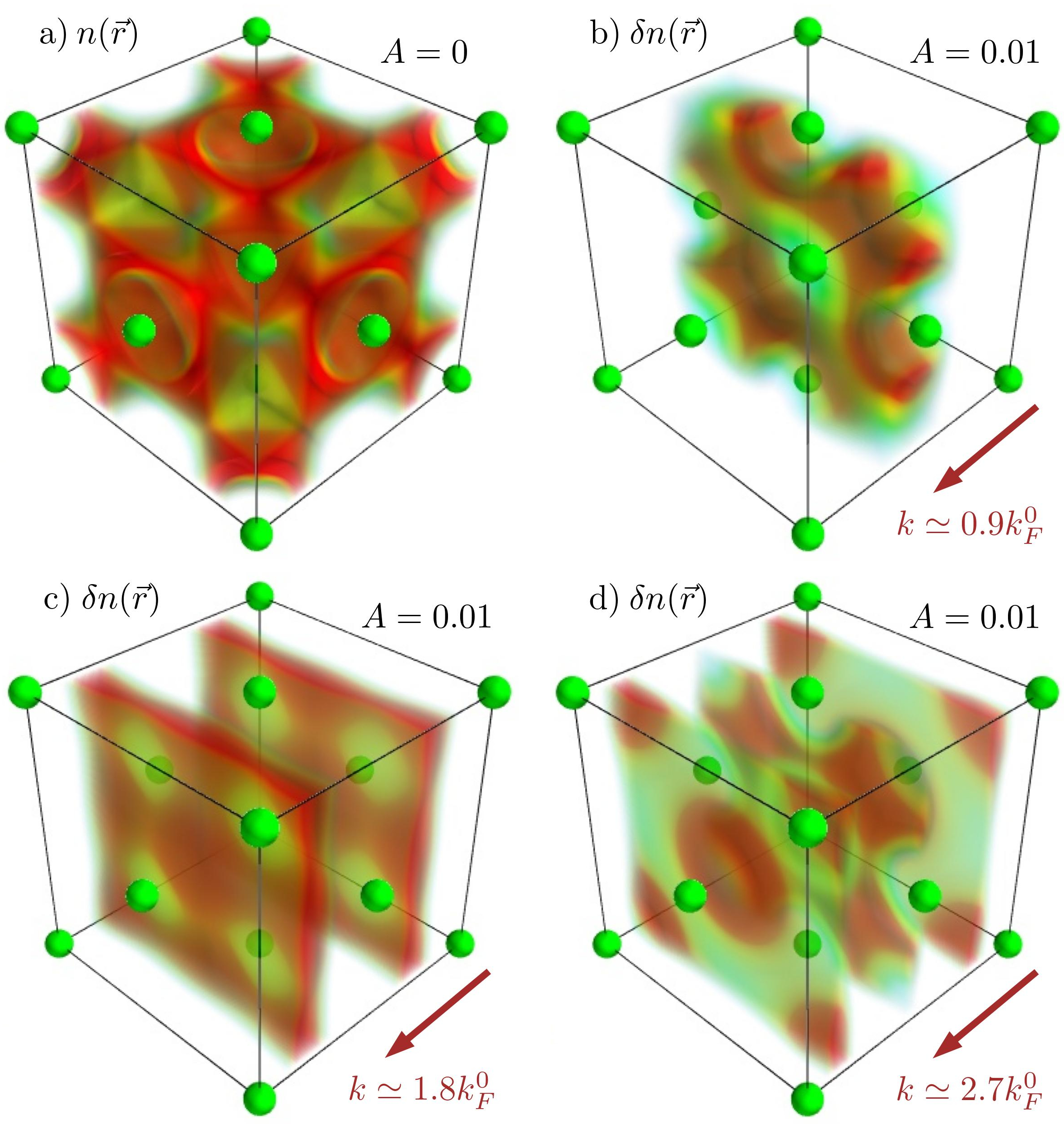}
\caption{\label{fig:3D_Al} Bulk electron density of Al with an fcc structure.
Volume rendering is used to visualize a) the electron density for  the unperturbed case ($A=0$), and the density perturbation due to an external harmonic perturbation with the  amplitude $A=0.01$ and wavenumbers b) $k\simeq 0.9k_F^0$, c) $k\simeq 1.8k_F^0$, and d)  $k\simeq 2.7k_F^0$. Al atoms are shown as green spheres. 
}
\end{figure}

\begin{figure}\centering
\includegraphics[width=0.48\textwidth]{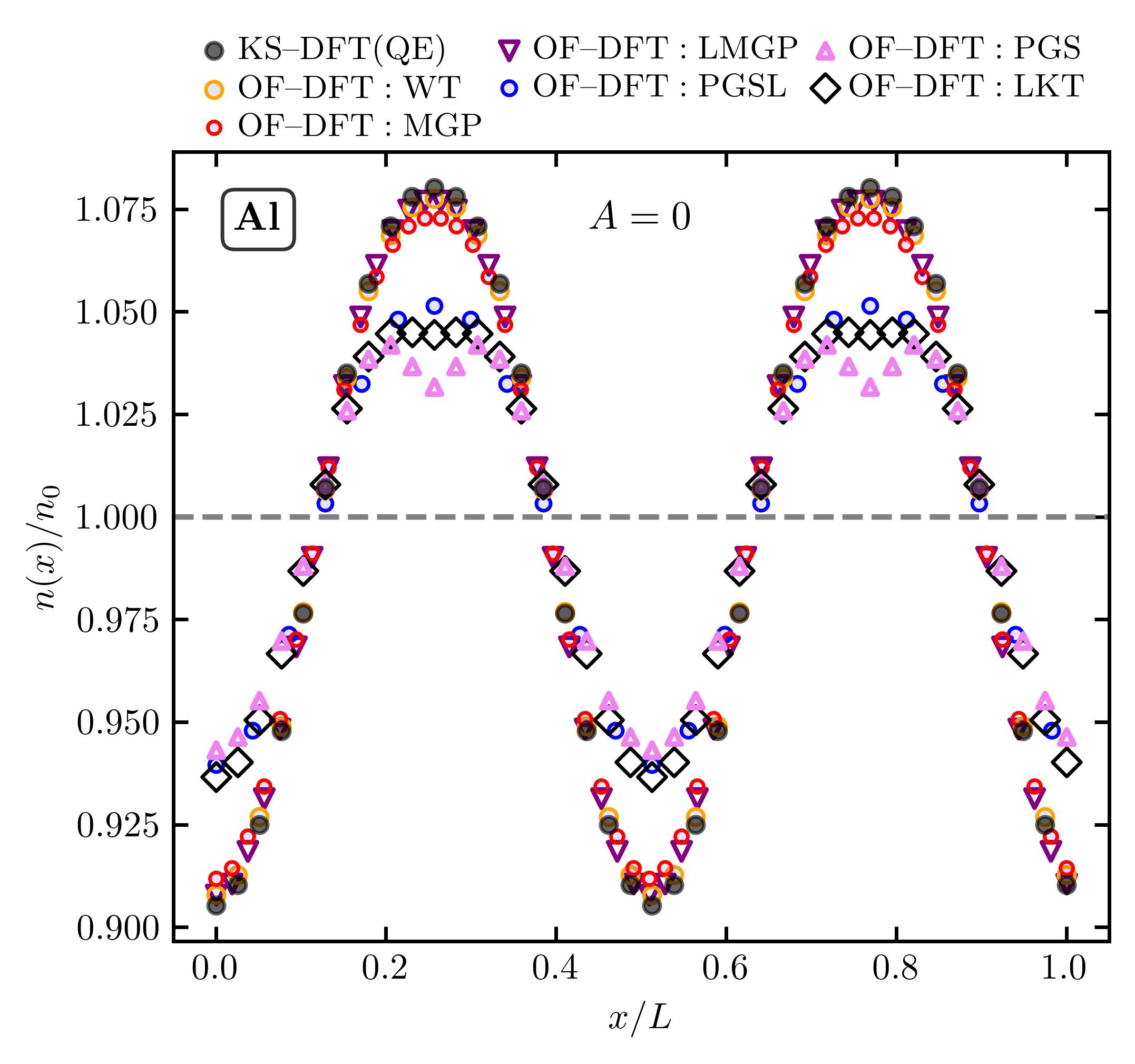}
\caption{\label{fig:Al_A0} Profile of the density projection $n(x)/n_0$ on the x-axis in units of the mean valence electron density ($n_0\simeq 1.81\times 10^{23}~{\rm cm^{-3}}$).  The results are computed for the primitive unit cell of fcc Al using OF-DFT with different KE functionals and KS-DFT; an LDA XC functional has been used for all calculations.
}
\end{figure}

\begin{figure}\centering
\includegraphics[width=0.48\textwidth]{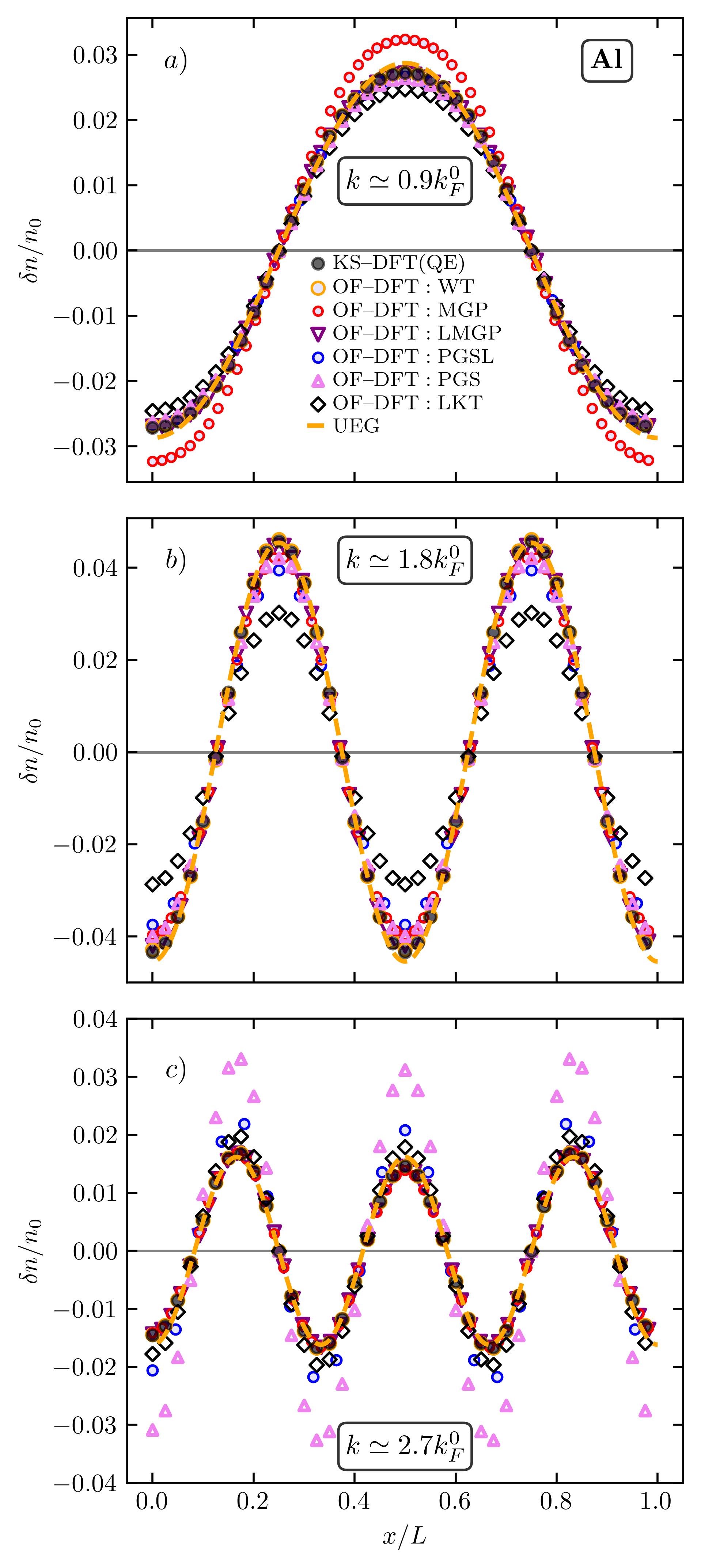}
\caption{\label{fig:Al_A0_01} Projection of  the density perturbation  profile $\delta n(x)/n_0$ on the x-axis in units of the mean valence electron density ($n_0\simeq 1.81\times 10^{23}~{\rm cm^{-3}}$).  The results are computed for the primitive unit cell of fcc Al using OF-DFT with different KE functionals and KS-DFT; an LDA XC functional has been used for all calculations. The density perturbation is induced by an external harmonic field with the  amplitude $A=0.01$ and wavenumbers a) $k\simeq 0.9k_F^0$, b) $k\simeq 1.8k_F^0$, and c)  $k\simeq 2.7k_F^0$.
}
\end{figure}

\subsection{Bulk Aluminum with fcc lattice}

As an example for the application of our approach to a real system, 
we next consider bulk Al with an fcc structure.
In general, the electronic density in materials is inhomogeneous. 
This is illustrated in Fig. \ref{fig:3D_Al}a), where we show the electron density in the unperturbed case using volume rendering. The results presented in Fig. \ref{fig:3D_Al} are computed using KS-DFT with an LDA \cite{LDA_PW} XC functional. Additionally, we show the density perturbations created by an external harmonic field with the amplitude $A=0.01$ and the wavenumbers b) $k\simeq 0.9k_F^0$, c) $k\simeq 1.8k_F^0$, and d) $k\simeq 2.7k_F^0$. In contrast to the UEG case, one can see from Fig. \ref{fig:3D_Al} that the density perturbation is also inhomogeneous in the  direction perpendicular to the external harmonic perturbation wave vector (which is indicated in Fig. \ref{fig:3D_Al} by an arrow). In order to quantitatively analyze the difference between OF-DFT results and KS-DFT results, we consider the projection of the density and density perturbation onto the x-axis along which the harmonic perturbation is directed. 

In Fig. \ref{fig:Al_A0}, we present the corresponding results for the density projection along the x-axis computed using OF-DFT with different KE functionals and using KS-DFT. We used the same LDA XC functional in all calculations. 
The density value is shown in the units of the mean density of the valence electrons $n_0\simeq 1.81\times 10^{23}~{\rm cm^{-3}}$. 

% From Fig. \ref{fig:Al_A0}, we note that the characteristic density inhomogeneity wavelength in the unit cell is about $\lambda \approx L/2$ , with $L=4.05~{\rm \AA} $ being the used fcc lattice parameter.

Examining the data in  Fig. \ref{fig:Al_A0}, we see that a meaningful analysis of the OF-DFT results can be performed by looking at $\Delta n=n-n_0$, which is the deviation of the density profile from the mean electron density $n_0$; the latter is depicted as the dashed horizontal line in Fig. \ref{fig:Al_A0}. First, we note that, in the case of the KS-DFT data, $\Delta n_{\rm KS-DFT}$ has the largest value about $0.075~n_0$, i.e., about $7.5\%$ of the mean density. As expected for the valence electrons of metals, this indicates that the valence electrons are weakly perturbed by the ions in the bulk region of Al with an fcc lattice. We note that the largest deviation amplitude in the density accumulation region is smaller than the largest density deviation magnitude in the density depletion region by about $17\%$. This  means that the density inhomogeneity  is not symmetric with respect to the positive and negative deviations from the mean density. 
Comparing  the $\Delta n$ data computed from the OF-DFT simulations with the results from the KS-DFT simulations, we find that the  WT and LMGP based results  closely reproduce the KS-DFT data, with a maximum relative error of about $3\%$. The MGP data exhibit a larger disagreement with the KS-DFT data  with  a maximum relative error in $\Delta n$ of about $9\%$.
In the case of the PGSL KE functional, the largest relative errors in $\Delta n$ are about $35\%$. 
For the LKT and the PGS KE functionals, we observe the largest relative errors in $\Delta n$ about  $44\%$  and $47\%$, respectively. 
Therefore, as one might expect, the semi-local GGA  KE functionals are  less accurate for the  OF-DFT calculations of the density compared to other considered fully nonlocal KE functionals in this case. The disagreement between OF-DFT and KS-DFT results are most pronounced around the maxima and minima of the density. We do not consider the effect of these observations on the calculation of other bulk properties since the effect of density errors can be masked or canceled to a certain degree by other effects, e.g., the behavior of the total energy~\cite{ke_test_jctc}. We next proceed to the analysis of the results for the density response to the external harmonic perturbation computed using OF-DFT and KS-DFT.

In Fig. \ref{fig:Al_A0_01}, we show the density perturbation $\delta n=n_{\rm A}-n_{\rm A=0}$ induced by an external harmonic potential with the  amplitude $A=0.01$ and wavenumbers a) $k\simeq 0.9k_F^0$, b) $k\simeq 1.8k_F^0$, and c)  $k\simeq 2.7k_F^0$. Additionally to the DFT data, we provide the density perturbation values computed using the density response function of the UEG according to Eq. (\ref{eq:chi_lda}) at $n_0\simeq 1.81\times 10^{23}~{\rm cm^{-3}}$ (dashed line). Comparing the KS-DFT results with the UEG model results, we see that  the UEG model is able to reproduce the KS-DFT data at all considered wavenumbers despite of the microscopic inhomogeneities in both the unperturbed  and perturbed density distributions (as illustrated in Fig. \ref{fig:3D_Al}). To understand this observation, we note that the used harmonic perturbation can be considered as macroscopic since it acts at all space points and the reaction of the system to this perturbation observed in Fig. \ref{fig:Al_A0_01} is a macroscopic density response since we have performed averaging over the density values in the direction perpendicular to the perturbation wavevector. The thus obtained macroscopic static density response of the valence electrons in metallic Al is accurately described by the UEG model. Although it is considered to be common knowledge, as far as we know, this had not been demonstrated quantitatively in this way. This example  demonstrates also how the direct perturbation approach can be used to acquire a physical picture of electronic properties in real materials.  

 \begin{figure}
\includegraphics[width=0.48\textwidth]{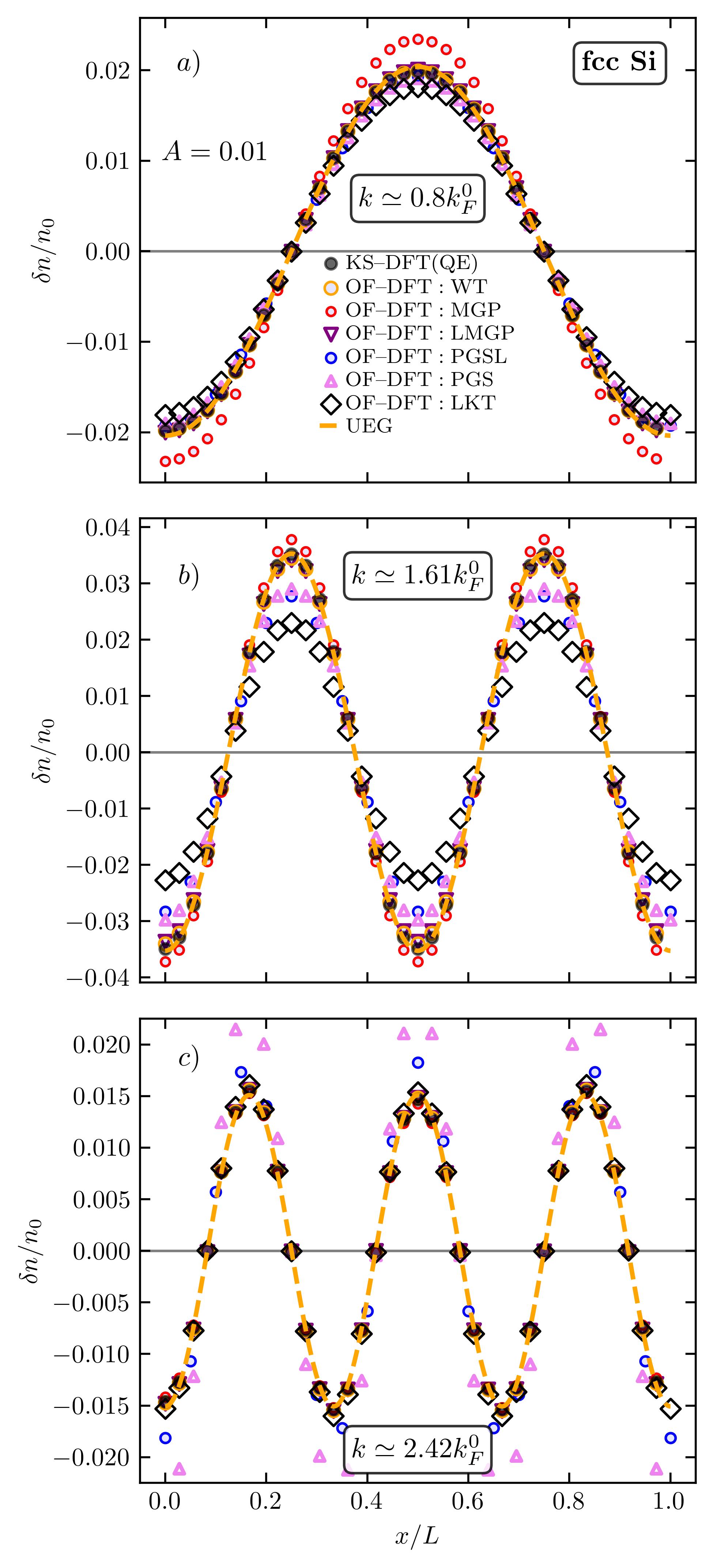}
\caption{\label{fig:fcc_Si_A0_01} Density response $\delta n(x)/n_0$  in units of the mean valence electron density ($n_0\simeq  2.82\times 10^{23}~{\rm cm^{-3}}$).  The results are computed for the primitive unit cell of fcc Si using OF-DFT with different KE functionals and KS-DFT; an LDA XC functional has been used for all calculations. The density perturbation is induced by an external harmonic field with the  amplitude $A=0.01$ and wavenumbers a) $k\simeq 0.8k_F^0$, b) $k\simeq 1.61k_F^0$, and c)  $k\simeq 2.42k_F^0$.
}
\end{figure} 

\begin{figure}
\includegraphics[width=0.48\textwidth]{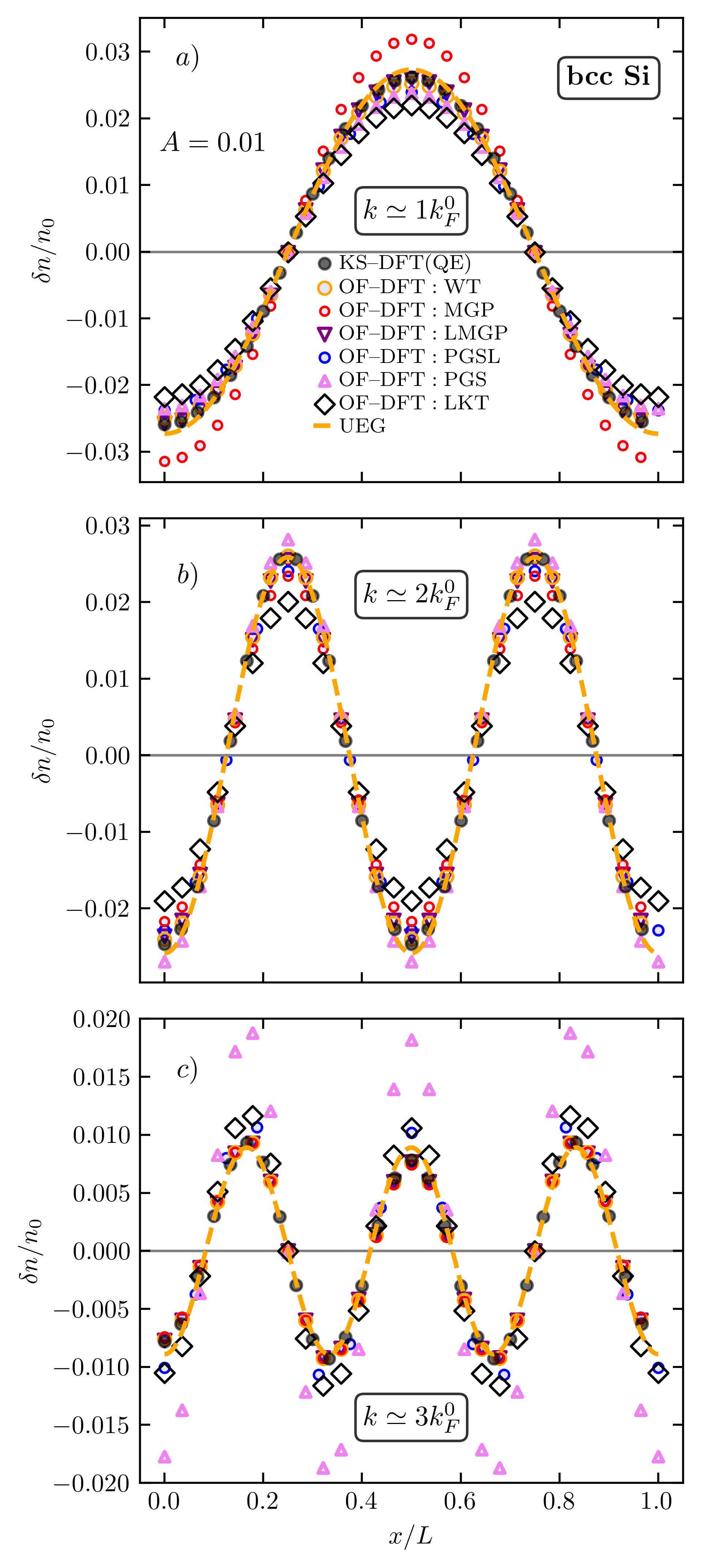}
\caption{\label{fig:bcc_Si_A0_01} Density response $\delta n(x)/n_0$  in units of the mean valence electron density ($n_0\simeq  2.75\times 10^{23}~{\rm cm^{-3}}$).  The results are computed for the primitive unit cell of bcc Si using OF-DFT with different KE functionals and KS-DFT; an LDA XC functional has been used for all calculations. The density perturbation is induced by an external harmonic field with the  amplitude $A=0.01$ and wavenumbers a) $k\simeq 1k_F^0$, b) $k\simeq 2k_F^0$, and c)  $k\simeq 3k_F^0$.
}
\end{figure}

Let us now summarize the performance of the considered KE functionals within OF-DFT for the description of the density perturbations presented in  Fig. \ref{fig:Al_A0_01}. Despite being generated on top of an inhomogeneous density,
the quality of the density response from OF-DFT relative to the KS-DFT results correlates with the observations for the UEG discussed in Sec. \ref{ss:results_A0_01}. The WT and LMGP based results closely reproduce the KS-DFT data.
In spite of the inaccurate unperturbed density values, the PGSL based results for the density perturbation by the weak external harmonic field  are in good agreement with  the KS-DFT data at $k=k_{\rm min}\simeq 0.9k_F^0$,  $k=2k_{\rm min}\simeq 1.8k_F^0$, and  $k=3k_{\rm min}\simeq 2.7k_F^0$ (being slightly worse for the latter). This can be due to the aforementioned effect of the averaging of the  density inhomogeneities and due to the small magnitudes of these density inhomogeneities compared to the mean density value.
The PGS data is similar to that of PGSL at $k=k_{\rm min}$ and  $k=2k_{\rm min}$, but is not able to adequately describe the density perturbation at  $k=3k_{\rm min}$. 
In the case of the MGP functional, we see from  Fig. \ref{fig:Al_A0_01} that the corresponding results have large deviations from the KS-DFT data at wavenumbers $k\simeq 0.9k_F^0$ and $k=3\simeq 2.7k_F^0$, and are in close agreement with the KS-DFT data  at $k\simeq 1.8k_F^0$. 
The LKT  based results significantly underestimate the density perturbation values at $k=2k_{\rm min}$ and are in rather close agreement with the  KS-DFT data at $k=k_{\rm min}$ and  $k=3k_{\rm min}$. This is consistent  with the pattern we have observed considering the density response of the UEG  in Sec. \ref{ss:results_A0_01}. 

\subsection{Bulk Silicon with fcc, bcc, and cd lattice}

Next, we consider bulk Si with an fcc lattice, body-centered cubic (bcc) lattice, and semiconducting crystal diamond (cd) phase.
In Fig. \ref{fig:fcc_Si_A0_01} and Fig. \ref{fig:bcc_Si_A0_01}, we show the density perturbation values at $k=k_{\rm min}=2\pi/L$,  $k=2k_{\rm min}$, and $k=3k_{\rm min}$, where $L$ is the corresponding  length of the unit cell. We set the XC functional to LDA \cite{LDA_PW} in both KS-DFT and OF-DFT simulations. For the fcc lattice, we have $k_{\rm min}\simeq 0.8k_F^0$ and for the bcc lattice we have $k_{\rm min}\simeq 1k_F^0$.
Similar to the fcc Al case and for the same reasons,  we see that the density perturbations in the fcc and bcc Si are accurately described by the UEG model. Correspondingly, the WT functional based OF-DFT data is in close agreement with the KS-DFT data for both fcc and bcc structures.
Similarly to the case of the fcc Al, the performance of the other considered KE functionals strongly correlates with their accuracy for the density response in the limit of  the UEG at relevant wavenumbers. 

\begin{figure}
\includegraphics[width=0.48\textwidth]{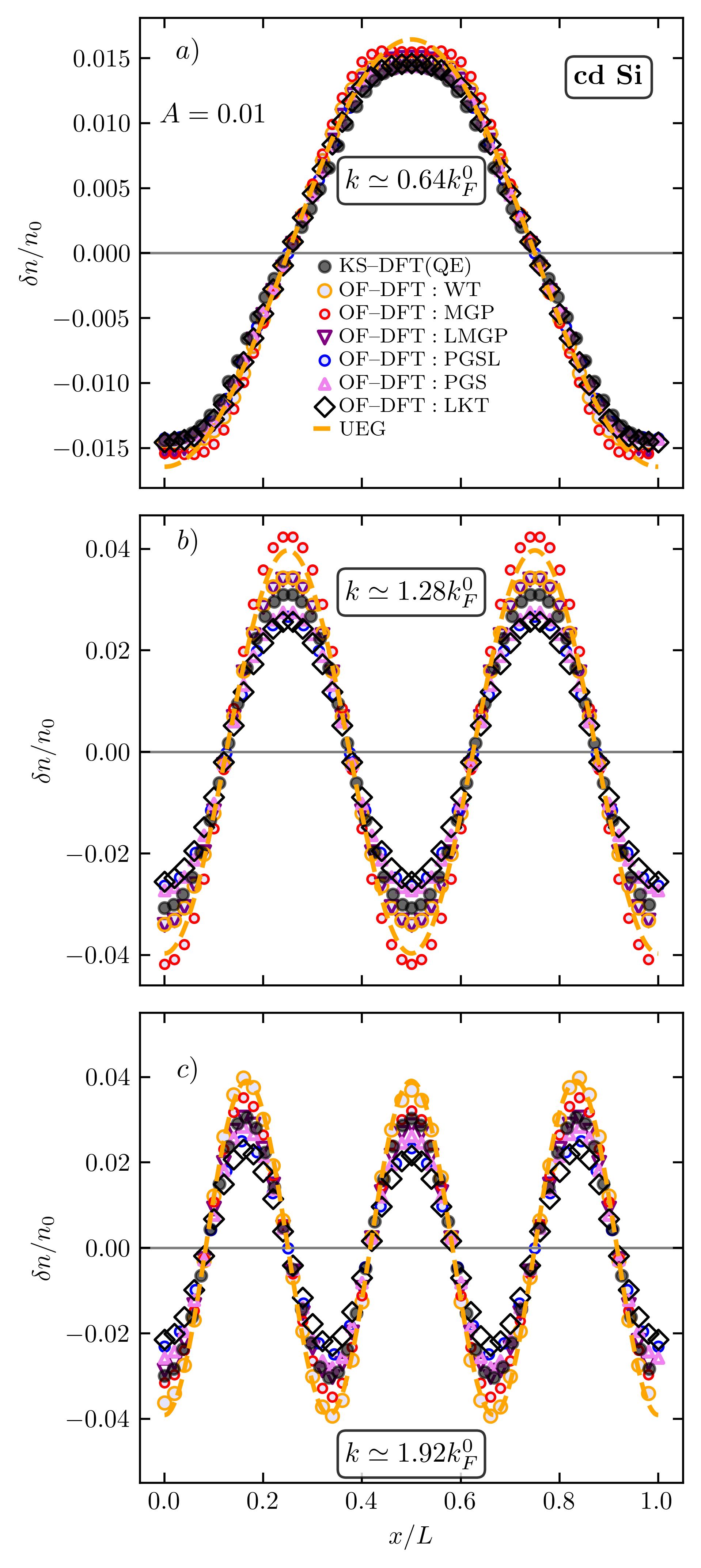}
\caption{\label{fig:cd_Si_A0_01} Density response $\delta n(x)/n_0$  in units of the mean valence electron density ($n_0\simeq  2.03\times 10^{23}~{\rm cm^{-3}}$).  The results are computed for the primitive unit cell of cd Si using OF-DFT with different KE functionals and KS-DFT; an LDA XC functional has been used for all calculations. The density perturbation is induced by an external harmonic field with the  amplitude $A=0.01$ and wavenumbers a) $k\simeq 0.64k_F^0$, b) $k\simeq 1.28k_F^0$, and c)  $k\simeq 1.92k_F^0$.
}
\end{figure}

In the case of the cd Si phase, we see from Fig. \ref{fig:cd_Si_A0_01} that the UEG model is less accurate. % for the description of the KS-DFT data. 
This is expected since cd Si is a semiconductor.
The disagreements between the UEG model and KS-DFT results are particularly pronounced at $k=2k_{\rm min}$ and $k=3k_{\rm min}$ (for cd Si we have $k_{\rm min}\simeq 0.64k_F^0$).
Despite being based on the UEG model (in the sense of Eq. (\ref{eq:start})), the WT, LMGP, PGSL, and PGS KE functionals (used in the OF-DFT simulations) provide better agreement with the KS-DFT data than the UEG model. This demonstrates the versatility of these functionals for the description of bulk metals and semiconductors \cite{ke_test_jctc, MGP, PGSL,shao2021efficient,huang2010nonlocal,LMGP}. Additionally, we observe that the LKT and PGS KE functionals provide results similar to each other for cd Si.

%%%%%%%%%%%%%%%%%%%%%%%%%%%%%%%%%%%%%%%%%%%%%%%%%%%%%%%%%%%%%%%%%%%%%%%%%%%%%%%%
% Discussion
%%%%%%%%%%%%%%%%%%%%%%%%%%%%%%%%%%%%%%%%%%%%%%%%%%%%%%%%%%%%%%%%%%%%%%%%%%%%%%%%
\section{Conclusions and Outlook}\label{s:end}

We have demonstrated the application of the direct perturbation approach for the test and analysis of various KE functionals for OF-DFT calculations.
By measuring the density response of the UEG to an external harmonic perturbation, we are able to crosscheck whether a KE functional satisfies the constraint (\ref{eq:start}). We emphasise that this tool is independent from the analytical reasoning used to construct a given KE functional. As a demonstration, we analyzed the density response generated by OF-DFT simulations using the nonlocal WT, MGP, LWT, LMGP functionals,  a Laplacian-meta-GGA level PGSL functional, and GGA level LKT and PGS functionals.
The WT, MGP, LWT, LMGP  functionals were built utilizing the constraint (\ref{eq:start}) for all wavenumbers and the PGSL (PGS) is designed to respect the constraint (\ref{eq:start})  at $k<2k_F^0$ ($k<k_F^0$). 
However, using the direct perturbation approach, we found that the MGP KE functional violates the constraint (\ref{eq:start}) at intermediate  wavenumbers $0.2\lesssim k/k_F^0 \lesssim 2.5$.
This illustrates the utility of the direct perturbation approach for testing KE functionals.

Going beyond the UEG limit, we analyzed the results computed using the considered KE functionals for the harmonically perturbed inhomogeneous electron gas over a wide range of wavenumbers and density perturbation degrees.
We found a strong correlation between the performance of the KE functionals in the UEG limit and in the strongly inhomogeneous case. This empirically demonstrates the importance of the constraint (\ref{eq:start}) for the construction of accurate KE functionals.

Furthermore, for the example of the PGSL and PGS KE functionals, we numerically validated the mapping of $s$ and $q$---constructed using density gradients---on $k/(2k_F^0)$ and $k^2/(2k_F^0)^2$ in the long wavelength expansion of the inverse density response function for the construction of KE functionals. This is an important result since such a mapping is also used for the construction of XC functionals using the local field correction (XC kernel) of the UEG (e.g., see Ref. \cite{Moldabekov_non_empirical_hybrid}).

The application of the direct perturbation approach for the analysis of the density response properties and performance of the KE functionals for real materials are demonstrated for the example of Al with an fcc lattice structure and Si with fcc, bcc, and cd phases. We demonstrated that the comparison of KS-DFT data for the density perturbation to the UEG model allows one to understand the role of the microscopic density inhomogeneity (induced by ions) with respect to the density response properties of the valence electrons. Despite of the microscopic inhomogeneities present in Al and in Si with fcc and bcc structures, the macroscopic density response of the valence electrons is accurately described by the UEG model. 
As the result, the quality of the considered KE functionals for the description of the density perturbation in fcc Al, fcc Si, and in bcc Si is similar to that  for UEG case. 
% \mp{
Despite being based on the UEG model, the WT and  LMGP  nonlocal functionals provide an adequate description of the density response of cd Si state. For this case,  the PGS and LKT  have similar quality and the PGSL performs best among considered semi-local functionals.  
% }

% \mp{I do not agree with this conclusion. It seems to me that the nonlocal functionals, with exception of MGP, perform very well for realistic systems. Instead the semimlocal functionals show a strong deviation (3-fold bigger than nonlocal functionals). PGS and LKT seem similar and PGSL is the best performer among them.}

In summary, the results clearly show that the quality of the static response functions of periodic bulk systems (showed for the example of the UEG, fcc Al, fcc Si, and bcc Si) require KE functionals that specifically encode UEG response behavior in the limiting case of constant ground state densities. While such a requirement is not important for perturbations with small wavevectors, it is crucial for high-$k$ perturbations with repercussions beyond ground-state OF-DFT. 

When modeling optical and ultrafast electronic properties of materials, time-dependent OF-DFT simulations of bulk systems, therefore, will need to move away from semilocal KE functionals and instead employ fully nonlocal functionals to be able to provide physical results for high-$k$ perturbations for periodic solids.

For modeling warm dense matter, Graziani \textit{et al} \cite{Graziani_CPP_2022} have recently pointed out  the possible importance of quantum nonlocality effects for the description of the shock propagation.
Preliminary results computed using a vW functional based quantum KE potential indicate that the induced density change at a shock front can reach about $0.5n_0$ for wavenumbers $k\lesssim k_F^0$.
Taking into account these findings and the result of the present analysis, we suggest that the potential generated by a semilocal (e.g., PGSL and PGS) KE functional can be used for a more reliable investigation of the impact of quantum nonlocality on the shock propagation in warm dense matter with strongly degenerate electrons. We single PGSL and PGS KE functionals due to their relatively simple form compared to the fully nonlocal KE functionals. This is advantageous for the implementation into existing hydrodynamics codes for which it is important to minimise a computational overhead due to calculation of the force field. Additionally,  semilocal KE functionals such as PGSL 
% and PGS \mp{I would only keep PGSL here}
are able to provide correct density response properties of electrons at $k\lesssim k_F^0$, which is important for the adequate description of the ion screening in warm dense matter \cite{ Moldabekov_pop2015, zhandos_cpp21, zhandos_cpp17}.  
We note that the finite temperature LKT functional \cite{PhysRevB.101.075116} and finite temperature WT functional \cite{Sjostrom_PRB2013} can be used to extend the present analysis into the WDM regime with partially degenerate electrons, and to the simulation of shock propagation at high temperatures relevant for inertial confinement fusion experiments. 

In this work, we considered the KE functionals constructed using the Lindhard density response function.
There are other types of the density response function that can be used for the construction of a KE functional, such as the quadratic density response function of the ideal electrons gas \cite{PhysRevB.100.125106, PhysRevB.100.125107}. Since the analytical forms of these density response functions are known, the presented harmonic perturbation based approach can be used as an independent tool for testing the correctness of the implementation of the KE functionals built using any of these density response functions.

Finally, we also note that a more detailed study of the response of semiconductors is the subject of future work, where in addition to the nonlocal KE functionals with density-independent kernel considered here, one can employ KE functionals based on the jellium-with-gap model \cite{PhysRevB.95.115153}.

% Acknowledgments
%%%%%%%%%%%%%%%%%%%%%%%%%%%%%%%%%%%%%%%%%%%%%%%%%%%%%%%%%%%%%%%%%%%%%%%%%%%%%%%%
\section*{Acknowledgments}
This work was funded by the Center for Advanced Systems Understanding (CASUS) which is financed by Germany’s Federal Ministry of Education and Research (BMBF) and by the Saxon state government out of the State budget approved by the Saxon State Parliament. This work has received funding from the European Research Council (ERC) under the European Union’s Horizon 2022 research and innovation programme
(Grant agreement No. 101076233, "PREXTREME"). 
Views and opinions expressed are however those of the authors only and do not necessarily reflect those of the European Union or the European Research Council Executive Agency. Neither the European Union nor the granting authority can be held responsible for them.
We gratefully acknowledge computation time at the Norddeutscher Verbund f\"ur Hoch- und H\"ochstleistungsrechnen (HLRN) under grant shp00026, and on the Bull Cluster at the Center for Information Services and High Performance Computing (ZIH) at Technische Universit\"at Dresden. MP and XS acknowledge the National Science Foundation under Grants No.\ CHE-2154760, OAC-1931473.

%%%%%%%%%%%%%%%%%%%%%%%%%%%%%%%%%%%%%%%%%%%%%%%%%%%%%%%%%%%%%%%%%%%%%%%%%%%%%%%
% Appendix
%%%%%%%%%%%%%%%%%%%%%%%%%%%%%%%%%%%%%%%%%%%%%%%%%%%%%%%%%%%%%%%%%%%%%%%%%%%%%%%
\appendix

\section{The calculation of the density response function from the density perturbation values}

\begin{figure}
\includegraphics[width=0.48\textwidth]{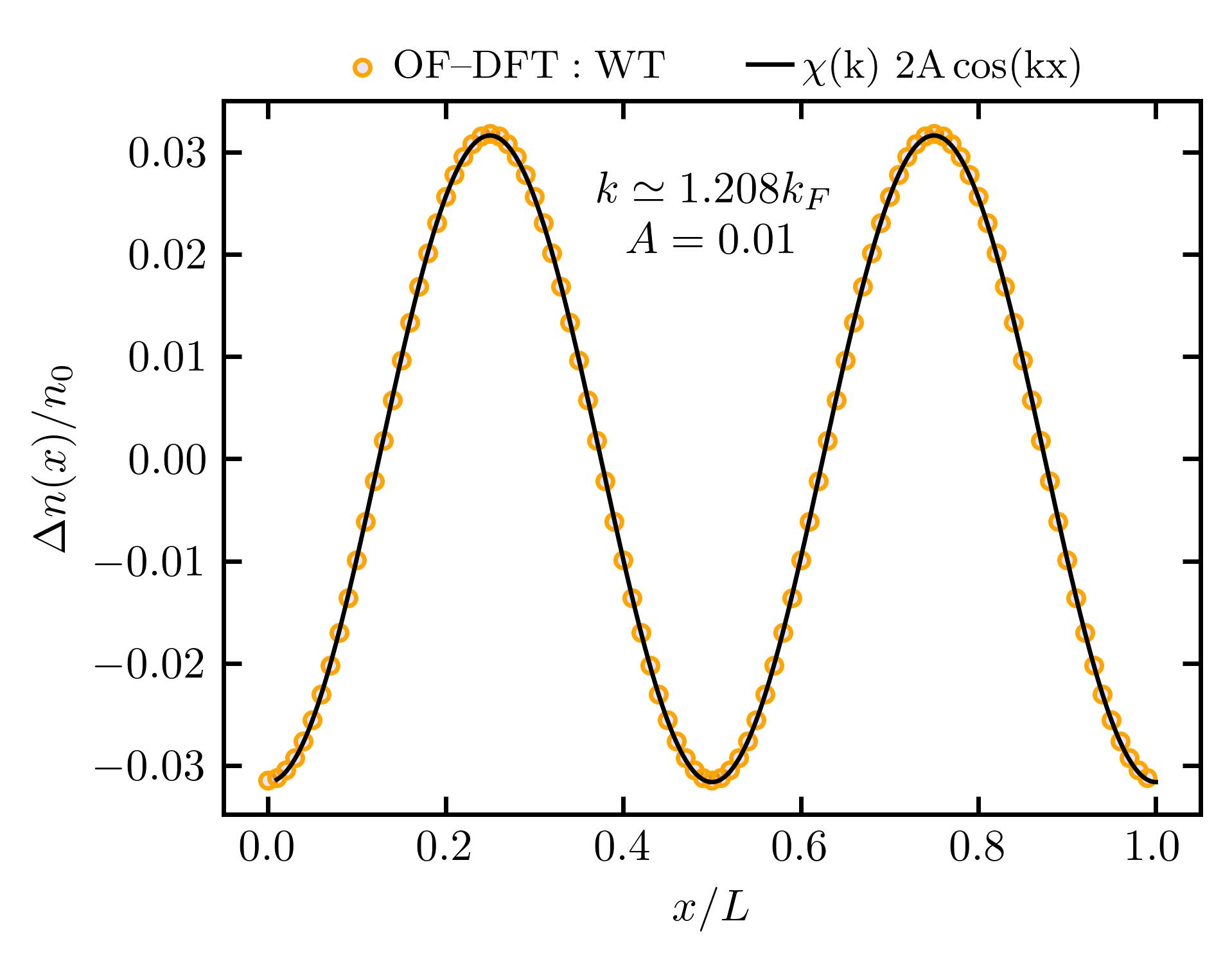}
\caption{\label{fig:A0_01_UEG} Density profile along the perturbation direction at $A=0.01$ and $k\simeq 1.208 k_F^0$. The  OF-DFT results are computed using the WT functional. The solid line represents the functional form $\chi(k) 2A\cos(kx)$ with $\chi(k)$ being computed from a fit to the OF-DFT results. 
}
\end{figure}

In Fig. \ref{fig:A0_01_UEG}, we show the density perturbation profile in the case of the ideal free electron gas with $A=0.01$ and $k\simeq 1.208 k_F^0$ (with perturbation direction along the x-axis).
The OF-DFT results are computed using the WT functional. The solid line follows the functional form $\chi(k) 2A\cos(kx)$, i.e,  Eq. (\ref{eq:chi}).
The value of $\chi(k)$ has been obtained by fitting the OF-DFT results using Eq. (\ref{eq:chi}).

%%%%%%%%%%%%%%%%%%%%%%%%%%%%%%%%%%%%%%%%%%%%%%%%%%%%%%%%%%%%%%%%%%%%%%%%%%%%%%%%
% METHODOLOGY
%%%%%%%%%%%%%%%%%%%%%%%%%%%%%%%%%%%%%%%%%%%%%%%%%%%%%%%%%%%%%%%%%%%%%%%%%%%%%%%%
% \section*{METHODOLOGY}

%%%%%%%%%%%%%%%%%%%%%%%%%%%%%%%%%%%%%%%%%%%%%%%%%%%%%%%%%%%%%%%%%%%%%%%%%%%%%%%%
% Bibliography
%%%%%%%%%%%%%%%%%%%%%%%%%%%%%%%%%%%%%%%%%%%%%%%%%%%%%%%%%%%%%%%%%%%%%%%%%%%%%%%%
\bibliography{bibliography.bib}

\end{document}